\documentclass[a4paper,12pt]{article}

\usepackage[pdftex]{graphicx}
\usepackage{subfigure}
\usepackage{float}
\newcommand{\be}{\begin{equation}}
\newcommand{\ee}{\end{equation}}

\newcommand{\bea}{\begin{eqnarray}}
\newcommand{\eea}{\end{eqnarray}}

\newcommand{\p}{\partial}

\newcommand{\nn}{\nonumber \\}
\newcommand{\f}{\frac}
\newcommand{\w}{\wedge}

\begin{document}
\thispagestyle{empty}

\begin{center} \noindent \Large \bf 
Transport properties of DBI action

\end{center}

\bigskip\bigskip\bigskip
\vskip 0.5cm
\begin{center}
{ \normalsize \bf   Shesansu Sekhar Pal}

\vskip 0.5 cm

 Department of Physics, Utkal University,  Bhubaneswar, 751004, India\\

\vskip 0.5 cm
\sf { shesansu${\frame{\shortstack{AT}}}$gmail.com }
\end{center}
\centerline{\bf \small Abstract}

We obtain  precise temperature dependence of three transport coefficients that resembles  strange metal:   (a) longitudinal electrical conductivity (b) longitudinal thermoelectric conductivity (c) longitudinal thermal conductivity in the absence of magnetic field. This is achieved through one parameter, namely, when the dynamical exponent, $z=4$. Moreover, in the presence of  magnetic field, we have cooked up a model where it exhibits the desired temperature dependence of  the four  transport quantities that resembles  strange metal. 

\newpage
\section{Introduction \& results}

\paragraph{Introduction:} It is known that the DC electrical conductivity of  Maxwell system in $d+1$ dimensional gravitational setup  with asymptotically AdS spacetime  at a temperature, $T_H$, goes as \cite{Kovtun:2008kx,Iqbal:2008by}
\be
\sigma^{xx}\sim T^{d-3}_H.
\ee

This particular temperature dependence of the electrical conductivity  gets modified 
in the presence of a dilaton dependent kinetic energy term of the gauge field.
 If the Maxwell action has the form
\be \label{maxwell_dilaton_action}
S_{Maxwell}=-\f{1}{4}\int d^{d+1}x  \sqrt{-g}~~Z(\phi) F_{MN}F^{MN}
\ee
then the electrical conductivity for such a system at the horizon\footnote{The radially conserved current allows one to evaluate the conductivity  for any choice of the radial coordinate \cite{Iqbal:2008by} including that at the horizon. } gets modified 
for a spacetime with dynamical exponent $z$ and scale violating parameter, $\theta=2\gamma$,  
which comes into picture particularly at IR. 
In such a scenario the conductivity\footnote{We provide a quick derivation of it in Appendix A without metric fluctuation.} without metric fluctuation at the horizon reads as
 \be\label{cond_maxwell}
 \sigma^{xx}(r_h)\sim Z(\phi(r_h)) ~~T^{\f{(d-3)(1-\gamma)}{z}}_H.
 \ee

We can convince  the structure of the conductivity by looking at the  conserved electric current associated to the action as written in eq(\ref{maxwell_dilaton_action})  and it reads as \cite{Iqbal:2008by}
\be\label{maxwell_current}
J^{\mu}_{Maxwell}=-Z(\phi) \sqrt{-g} F^{r\mu},
\ee

where $r$ is the holographic direction. Since it is radially conserved means it can be evaluated for any value of the radial coordinate which will give the same answer. We choose to evaluate it at the horizon. The above form of the conductivity follows by looking at the x-component of the current with an imposition of the in-falling boundary condition on the x-component of the gauge potential at the horizon. The gauge potential has the non-normalizable mode  which is essentially identified with an electric field \cite{Donos:2014cya}. For a recent review on the study of  transport  properties of Maxwell system, see \cite{Hartnoll:2016apf} and reference therein.

Moving over to the DBI case, it is known that the action is described by a square-root term as follows
\be
S_{DBI}=-T_b Z_1(\phi)\sqrt{-det\bigg( Z_2(\phi) g_{MN}+\lambda F_{MN}\bigg)}
\ee

The  conserved current that follows  takes the form
\be
J^{\mu}_{DBI}=T_b\f{Z_1(\phi) \sqrt{-det\bigg( Z_2(\phi) g_{MN}+\lambda F_{MN}\bigg)}}{2}\left[\bigg( Z_2(\phi) g+\lambda F\bigg)^{-1~ r\mu}-\bigg( Z_2(\phi) g+\lambda F\bigg)^{-1~\mu r}\right].
\ee

Using the fluctuation of the  gauge potential as written down in eq(\ref{flu_metric_gauge_withgout_b}) and evaluating the currents at the horizon along with the imposition of the in-falling boundary condition 
gives rise to the following expression of the conductivity\footnote{The expression written in eq(\ref{cond_dbi}) is exact provided there is no fluctuation in the metric.} in $3+1$ dimensional spcetime \cite{Karch:2007pd,Hartnoll:2009ns,Pal:2010sx,Pal:2012zn}
\bea\label{cond_dbi}
 \sigma^{xx}(r_h)&=&T_bZ_1(\phi(r_h))\sqrt{1+\f{\rho^2}{(h(r_h)Z_1(\phi(r_h))Z_2(\phi(r_h)))^2}}\nn&=&T_bZ_1(\phi(r_h))\sqrt{1+\f{\rho^2T^{-\f{4(1-\gamma)}{z}}_H}{(Z_1(\phi(r_h))Z_2(\phi(r_h)))^2}}\quad {\rm for}\quad  h(r_h)=r^{2(1-\gamma)}_h=T^{\f{2(1-\gamma)}{z}}_H,
\eea
where we have included a background charge density, $\rho$. In the absence of such a charge density, the electrical conductivity that follows  matches with that follows form Maxwell action. Such type of square-root form of the electrical conductivity is also shown in \cite{Kiritsis:2016cpm,Cremonini:2017qwq, Blake:2017qgd}. 

It is interesting to note that the electrical conductivity that follows  from Maxwell action  for unit coupling $Z(\phi(r_h))=1$ in  $3+1$ dimensional spcetime is independent of temperature, see eq(\ref{cond_maxwell}), whereas the   electrical conductivity that follows  from DBI action does depends on temperature even for unit couplings $Z_1(\phi(r_h))=1=Z_2(\phi(r_h))$, see eq(\ref{cond_dbi}). In the limit of vanishing charge density the electrical conductivity for the DBI system goes over to the Maxwell system for $d=3$.

The expression of the electrical conductivities as written down in eq(\ref{cond_maxwell}) and eq(\ref{cond_dbi}) does not break any translational symmetry of the system.  In  solids it is expected not to have such a symmetry. A recipe to calculate the transport quantities in the absence of such a symmetry are proposed in \cite{Donos:2014cya} and  \cite{Gouteraux:2014hca}. The parameter that breaks the symmetry is interpreted to be the dissipation parameter \cite{Donos:2014cya}.

\paragraph{Results:} In this paper, we consider the Einstein-DBI-dilaton-axion system and study the temperature dependence of the transport coefficients in $3+1$ dimensional spacetime at IR. The results of our study is as follows.

$\bullet$  We show that the form of the transport quantities at the horizon eq(\ref{transport}) remain same irrespective of the exact form of the $tt$ component of metric, $g_{tt}$, and the $rr$ component of the metric, $g_{rr}$,  as long as $g_{tt}$ admits a zero and $g_{rr}$ admits a pole at the horizon.

$\bullet$ We show that the dissipation parameter, $k$, disappears completely from the result of the transport quantities at IR. Generically,  the dissipation parameter comes through   the axionic field. The presence of such fields makes the breakdown of the  translational symmetry in the system. In fact,  the above claim of the disappearance of the dissipation parameter from the result of the transport quantities works  in the presence of a non-trivial dilaton dependent kinetic energy term.
However, if the  kinetic energy term of the axionic fields are  canonically normalized i.e., do not contain any dilaton dependent term then the transport quantities depends on the dissipation parameter, $k$. This typically happens at UV \cite{Pal:2019bfw}.
This we demonstrate by going through an example of DBI system with gravity, dilaton gauge field and axionic fields. The explicit solution of such a system is generated at IR. It is described by a dynamical exponent, $z$,  and  the hyper scale violating (HSV) parameter, $\theta$ \cite{Huijse:2011ef}. 

If we look at the general expression of the transport coefficients as written down in eq(\ref{gen_transport}) and eq(\ref{gen_kappa}) then it follows that the dissipation parameter, $k$, and the value of the dilaton dependent coupling term,  $\psi$, evaluated at the horizon comes in a combination: $k^2~~ \psi(\phi(r_h))$. From the Lifshitz solution and hyper scale violating solution of the system under study, it follows that such a combination is independent of the dissipation parameter. This is simply due to  the coupling term $\psi$ in these cases goes as $1/k^2$.
Hence the combination $k^2~~ \psi(\phi(r_h))$ is independent of the dissipation parameter. The results of the transport quantities are reported in eq(\ref{lifshitz_transport}),  eq(\ref{hsv_transport_without_b}) and eq(\ref{hsv_transport}) for Lifshitz solution and hyper scale violating solution, respectively. However, at UV if we fix $\psi$ to be identity then the combination $k^2~~ \psi(\phi(r_h))$ goes  quadratic in the dissipation parameter.

\begin{table}
	\begin{center}
		\begin{tabular}{ |l | p{3cm} | p{2cm} |}
			\hline
			Transport quantities in Einstein-DBI system  & Strange metal & HSV \\ \hline\hline
			Longitudinal electrical conductivity, $\sigma_{xx}=\sigma_{yy}$ &$ T^{-1}_H$& $ T^{-\f{4-\theta}{z}}_H$\\\hline
			Longitudinal thermoelectric conductivity, $\alpha_{xx}=\alpha_{yy}$ &$\rho T^{-\f{1}{2}}_H$&$ \rho T^{-\f{2}{z}}_H$\\\hline
			Longitudinal thermal conductivity, $\kappa_{xx}=\kappa_{yy}$ &$T_H$& $T^{\f{z-\theta}{z}}_H$\\\hline
		Seebeck coefficient, $\alpha_{xx}/\sigma_{xx}$&$\rho T^{\f{1}{2}}_H$&$\rho T^{\f{(2-\theta)}{z}}_H$\\\hline
		\end{tabular}
	\end{center}\caption{\label{table}The precise temperature dependence of the transport  quantities  in $3+1$ spacetime dimension without magnetic field. $z$ is the dynamical exponent and $\theta=2\gamma$ is the scale violating parameter.}
\end{table}

Recently, there has been several attempts to understand the transport properties of the strange metal from holographic point of view using scaling arguments \cite{Khveshchenko:2014nka,Karch:2014mba,Hartnoll:2015sea}. In \cite{Hartnoll:2015sea}, the authors upon using the scaling arguments claims to have reproduced five distinct transport properties with only two non-zero exponents: dynamical exponent, $z=4/3$ and anomalous scaling exponent, $\Phi=-2/3$. The latter, anomalous scaling dimension, is associated with the charge density operator.  It is interesting to note that the hyperscaling violation exponent does not play any  role at all  in the determination of the temperature dependence of those transport coefficients. It should be mentioned {\it en passant} that the effect of charge density  is not considered at all in their study of the determination of the transport quantities.

$\bullet$ In our explicit construction of the model  without magnetic field, we show  in $3+1$ spaceime dimension the temperature dependence of three transport coefficients, namely, (a) longitudinal electrical conductivity (b) longitudinal thermal conductivity (c) ratio of thermo-electrical conductivity to the longitudinal electrical conductivity share the same temperature dependence as that of the non-Fermi Liquid or strange metal. The temperature dependence of these three transport coefficients are given in table(1) for any $z$ and $\theta\equiv 2\gamma.$ The precise temperature dependence of the transport coefficients for strange metal follows
upon setting  $z=4$ and $\gamma=0\equiv\theta/2$. One of the main findings of the paper is to generate the precise temperature dependence of these three transport quantities with one non-zero exponent, $z$.  

$\bullet$ As soon as we turn on the magnetic field the precise  temperature dependence of the transport coefficients becomes   very difficult to calculate, in general, because of the difficulties to find  real valued solutions. However, we found an exact real valued solution, which is a hyperscale violating solution with $\theta=4, ~z=$arbitrary. For this case, surprisingly, the electrical conductivity  does not show any temperature dependence. 

In order to have some non-trivial temperature dependence of the transport coefficients, we have made some specific choices of couplings and the spatial components of the metric. In particular, we choose the dilaton dependent kinetic energy term    of the axionic field  to be proportional to entropy density.  These choices  essentially  gives the desired temperature dependence of the four transport coefficients that resembles that of the strange metal. A comparison of these results are given in table (2).

\paragraph{Organization:} The paper is organized as follows: {\bf In section 2}, we shall present a scaling argument to find the  temperature dependence of the transport coefficients that resembles the strange metal. In particular, we have  included the effect of charge density in the longitudinal  thermoelectric conductivity, $\alpha_{xx}$. In the absence of magnetic field, this imposes a constraint on the dynamical exponent to be $z=4$, hyperscale violating exponent, $\theta=0$ and the anomalous scaling dimension of the charge density, $\Phi=-2$. The temperature dependence of the transport coefficients are also obtained by including both the charge density and the magnetic field. {\bf In section 3}, we setup the Einstein-DBI-dilaton-axion model and obtain the necessary equations that solve the equations of motion without magnetic field. We derive the currents and the transport quantities. {\bf In section 4}, we find the precise temperature of the transport coefficients by finding exact solutions of the model under study. {\bf In section 5}, we have included both the charge density and magnetic field and studied its effect on the transport coefficients. {\bf In section 6}, we provide a discussion on time reversal invariance and particle-hole symmetry. {\bf In Appendix A}, we give a quick derivation of the longitudinal electrical conductivity for Maxwell case.

\section{Scaling argument}

In the study of the  use of the scaling argument, it is typically the  physical quantities  are considered to have  mass dimension as follows.
 The time has the dimension
 \be
 [t]=-z
 \ee
This is mainly due to that the length has the mass dimension
\be
[x]=-1
\ee 
and the theory possess the dispersion relation: $\omega\sim k^z$. If the number of spatial dimensions in a given field theory is $d$ with the scaling violation exponent, $\theta$, then the entropy goes as $s\sim T^{\f{d-\theta}{z}}$ \cite{Huijse:2011ef}. It gives 
\be
[s]=d-\theta,
\ee
then it follows trivially that the energy and free energy has the following mass dimension
\be
[f]=[E]=d-\theta+z.
\ee
Introducing an exponent, $\Phi$, as the  anomalous scaling dimension associated to the charge density operator allows us to give the mass dimension to the charge density as
\be
[\rho_e]=d-\theta+\Phi.
\ee

The scalar potential and the vector potential has the dimension
\be
[A_t]=z-\Phi,\quad [{\vec A}]=1-\Phi
\ee  
Given such scaling dimensions, it follows naturally that the electric current density and heat current density has the following dimensions 
\be
[{\vec J_e}]= d-\theta+\Phi+z-1,\quad [{\vec Q}]= d-\theta+2z-1
\ee

The electrical conductivity, thermoelectric conductivity and the thermal conductivity has the following dimensions

\be
[\sigma]=d-\theta+2\Phi-2,\quad [\alpha]=d-\theta+\Phi-2,\quad [\kappa]=d-\theta+z-2.
\ee

Upon assuming  the electrical conductivity, thermo-electric conductivity and thermal conductivities are determined completely by  temperature and magnetic field   gives desired form of the electrical conductivity, Hall angle, Lorentz ratio, thermopower and magneto-resistance  etc. only  for $\theta=0,~~z=4/3$ and $\Phi=-2/3$ \cite{Hartnoll:2015sea}. 

It is reported in \cite{behnia} that the temperature dependence of  thermo-electric conductivity typically arises due to the presence of   non-zero charge density  in either the electric current or in the heat current. If we include the dependence of the  charge density linearly in the thermo-electric conductivity   suggests us to write the   conductivities as 
\be
\sigma_{xx}\sim T^{\f{d-\theta+2\Phi-2}{z}}_H,\quad \alpha_{xx}\sim \rho T^{-\f{2}{z}}_H, \quad \kappa_{xx}\sim T^{\f{d-\theta+z-2}{z}}_H
\ee

Demanding the linear temperature dependece of the resistivity for $d=2$ as required for strange metal gives 
\be
\sigma_{xx}\sim T^{-1}\quad  \Longrightarrow \quad 2\Phi-\theta=-z
\ee

Assuming the thermopower or the seebeck coefficient have the following form
\be
S\equiv \f{\alpha_{xx}}{\sigma_{xx}}=\rho T^{-\f{2\Phi-\theta+2}{z}}_H= \rho T^{\f{z-2}{z}}_H =\rho T^{\f{1}{2}}_H\quad  \Longrightarrow \quad  z=4.
\ee

Assuming the thermal conductivity has the linear temperature dependence gives
\be
\kappa_{xx}\sim T^{\f{z-\theta}{z}}_H=T\quad  \Longrightarrow \quad \theta=0.
\ee

Finally, the desired temperature dependence  of  various transport coefficients as required for strange metals gives  
\be
\theta=0,\quad \Phi=-2,\quad z=4.
\ee
This result is obtained without  including any magnetic field in the transport coefficients. It is interesting to note that the exponent $\Phi$ comes out to be negative this time as well.

 In  the presence of magnetic field, the temperature dependence of the transport quantities  can easily be computed and reads  as follows:
 
\bea
\sigma_{xx}&\sim& T^{\f{d-\theta+2\Phi-2}{z}}_H,\quad \sigma_{xy}\sim \rho B T^{\f{2\Phi-4}{z}}_H, \quad \alpha_{xx}\sim \rho T^{-\f{2}{z}}_H,\quad \alpha_{xy}\sim B T^{\f{d-\theta+2\Phi-4}{z}}_H, \nn \kappa_{xx}&\sim& T^{\f{d-\theta+z-2}{z}}_H,\quad \kappa_{xy}\sim \rho B  T^{\f{z-4}{z}}_H.
\eea

If we consider $d=2,~\theta=0,~ \Phi=-2,~ z=4$, then the transport quantities reads as

\bea\label{scaling_transport}
\sigma^{xx}&\sim& T^{-1}_H,\quad \sigma^{xy}\sim \rho BT^{-2}_H,\quad\kappa^{xx}\sim T_H,\quad\kappa^{xy}\sim\rho B T^{0}_H,\nn \alpha^{xx}&\sim&\rho T^{-\f{1}{2}}_H,\quad \alpha^{xy}\sim B T^{-3/2}_H,
\eea

Note, in contrast to \cite{Hartnoll:2015sea}, we have included the effect of charge density in the temperature dependence of the transport quantities. This is due to the suggestion in \cite{behnia}.  In fact,  the holographic calculation of the transport quantities do   generate such  dependence of  charge density very naturally. 
 
 Surprisingly, the scaling analysis does not give the desired temperature dependence (as seen for strange metals) of the transverse transport quantities like $\sigma_{xy},~\alpha_{xy},~\kappa_{xy}$ after the inclusion of the magnetic field. 
 This in fact allows us to do some model building as we have done in subsection 8.6.
 Interestingly, in  our holographic computation, we show that the desired temperature dependence of the transport quantities follows  without including the anomalous scaling dimension, $\Phi$, in the gauge potential.

 \paragraph{Summary:} In order to have the necessary temperature dependence of the transport quantities, the scaling analysis for $d=2$ with zero scale violation, $\theta=0$ gives a  relation between anomalous scaling dimension, $\Phi$, and the dynamical exponent, $z$, namely $z=-2\Phi$. Without the charge density in the transport quantities gives $z=4/3$ \cite{Karch:2014mba}, whereas with charge density, we have $z=4$.

\section{Model}
The action that we shall be considering contains metric, $g_{MN}$, gauge field, $A_M$ dilaton, $\phi$, and two axionic fields, $\chi_1,~\chi_2$. The action of the gauge field is through DBI action. This is mainly because we want to see the effect of the inclusion of the non-linear terms in the square of the field strength tensor on the transport quantities. 
\bea
S&=&\f{1}{2\kappa^2}\int d^{3+1} x\bigg[ \sqrt{-g}\Bigg({\cal R}-2\Lambda-\f{1}{2}(\p \phi)^2-V(\phi)-\f{W_1(\phi)}{2}(\p \chi_1)^2-\f{W_2(\phi)}{2}(\p \chi_2)^2\Bigg)\nn &-&T_b Z_1(\phi)\sqrt{-det\bigg( Z_2(\phi) g_{MN}+\lambda F_{MN}\bigg)}\Bigg],
\eea
which essentially describes the action of a space filling brane\footnote{For study of some interesting properties of such branes see \cite{Lee:2011qu,Pal:2017xpg,Pal:2019bfw,Gushterov:2018spg}.}.  
The equation of motion of the metric components that follows has the from 
\bea
&&{\cal R}_{MN}-\f{W_1(\phi)}{2}\p_M \chi_1\p_N \chi_1-\f{W_2(\phi)}{2}\p_M \chi_2\p_N \chi_2-\f{[V(\phi)+2\Lambda]}{2}g_{MN}-
\f{1}{2}\p_M\phi\p_N\phi\nn&&-\f{T_b}{8}\f{\sqrt{-det\bigg(g Z_2(\phi)+\lambda F\bigg)_{PS}}}{\sqrt{-g}}Z_1(\phi)Z_2(\phi)
\bigg[\bigg(gZ_2(\phi)+\lambda F\bigg)^{-1}+{\bigg(gZ_2(\phi)~ -\lambda F\bigg)}^{-1} \bigg]^{KL}\times\nn&&
\bigg[g_{MN}g_{KL}-2g_{MK}g_{NL} \bigg]=0\nn&&
\eea

and that of the gauge field is

\bea\label{gauge_field_eom}
&& \p_M\Bigg[ Z_1(\phi)\sqrt{-det\bigg(g Z_2(\phi)+\lambda F\bigg)_{PS}}\Bigg(\bigg(gZ_2(\phi)+\lambda F\bigg)^{-1}-{\bigg(gZ_2(\phi)-\lambda F\bigg)}^{-1}  \Bigg)^{MN}\Bigg]=0,\nn &{\rm where}&F_{MN}=\p_M A_N-\p_N A_M.
\eea

The equation of motion associated to the scalar field is
\bea\label{scalar_eom}
&&\p_{M}\bigg(\sqrt{-g}\p^M\phi \bigg)-\sqrt{-g}\bigg[\f{dV(\phi)}{d\phi}+\f{1}{2}\f{dW_1(\phi)}{d\phi}(\p \chi_1)^2+\f{1}{2}\f{dW_2(\phi)}{d\phi}(\p \chi_2)^2\bigg]\nn&&-T_b \f{dZ_1(\phi)}{d\phi}\sqrt{-det\bigg(g~ Z_2(\phi)+\lambda F\bigg)_{KL}}-\nn&&\f{T_b}{2}Z_1(\phi)
\sqrt{-det\bigg(g~ Z_2(\phi)+\lambda F\bigg)_{KL}}\f{dZ_2(\phi)}{d\phi}{\bigg(g~ Z_2(\phi)+\lambda F\bigg)}^{-1MN}g_{MN}=0.
\eea

The equation of motion for the axionic fields are
\be
\p_{M}\bigg(\sqrt{-g}W_1(\phi)\p^M\chi_1\bigg)=0,\quad \p_{M}\bigg(\sqrt{-g}W_2(\phi)\p^M\chi_2\bigg)=0
\ee
\subsection{Solution with dissipation}
\paragraph {Ansatz:}

\bea\label{ansatz_solution}
ds^2_{3+1}&=&-g_{tt}(r)dt^2+g_{rr}(r)dr^2+g_{xx}(r)(dx^2_1+dx^2_2),\quad \chi_1=k x_1,\quad \chi_2=k x_2\nn
F&=&A_t'(r) dr\w dt,\quad \phi=\phi(r)
\eea

With such an ansatz and setting $\lambda=1$, the  DBI gauge field can solved exactly in terms of the metric component as
\be
 A'_t=\rho
\f{Z_2\sqrt{g_{tt}g_{rr}}}{\sqrt{Z^2_1Z^2_2g^{2}_{xx}+\rho^2}},
\ee 
where $\rho$ is the charge density.

The equation of motion of the metric components can be expressed, explicitly,  as follows
\bea\label{detailed_metric_eom_dbi_gtt}
&&{\cal R}_{tt}+\f{(V+2\Lambda)}{2}g_{tt}+
\f{T_b}{2}Z^2_1Z^3_2\f{g_{tt}g_{xx} }{{\sqrt{\rho^2+ Z^2_1Z^2_2g^{2}_{xx}}}}=0,
\eea
\bea\label{detailed_metric_eom_dbi_dilaton_gxx}
&&{\cal R}_{ij}-\f{(V+2\Lambda)}{2}g_{xx}\delta_{ij}-\f{W_1}{2}k^2\delta_{ij}
-\f{T_b}{2}Z_2\delta_{ij} {\sqrt{\rho^2+Z^2_1Z^2_2g^{2}_{xx}}}=0,
\eea
\bea\label{detailed_metric_eom_dbi_dilaton_grr}
&&{\cal R}_{rr}-\f{(V+2\Lambda)}{2}g_{rr}-\f{\phi^{'2}}{2}-\f{T_b}{2} Z^2_1Z^3_2 \f{g_{rr}g_{xx}}{\sqrt{\rho^2+Z^2_1Z^2_2g^{2}_{xx}}}=0,
\eea
whereas the equation of motion associated to scalar field $\phi(r)$ becomes
\bea
&&\p_r\left(\f{g_{xx}\sqrt{g_{tt}}}{\sqrt{g_{rr}}}\phi'\right)-\sqrt{g_{tt}g_{rr}}
\left[ \f{dV(\phi)}{d\phi}+\f{k^2}{2}\left(\f{dW_1(\phi)}{d\phi}+\f{dW_2(\phi)}{d\phi}\right)\right]\nn&&+T_bZ_1Z_2\f{dZ_1(\phi)}{d\phi}\f{\sqrt{g_{tt}g_{rr}}Z^2_2 g^2_{xx}}{\sqrt{\rho^2+Z^2_1Z^2_2g^{2}_{xx}
		}}-T_b \f{dZ_2(\phi)}{d\phi}\sqrt{g_{tt}g_{rr}}
	\sqrt{\rho^2+Z^2_1Z^2_2g^{2}_{xx}}\nn&&-
		T_bZ^2_1Z^2_2\f{dZ_2(\phi)}{d\phi}\f{\sqrt{g_{tt}g_{rr}}Z^2_2 g^4_{xx}}{\sqrt{\rho^2+Z^2_1Z^2_2g^{2}_{xx}
		}}=0.
\eea

The equation of motion associated to axionic fields are satisfied automatically.
\paragraph{ Solution at UV: }For constant scalar field, $\phi(r)=\phi_0$, where $\phi_0$ is a constant.
One can easily check the following relation: eq(\ref{detailed_metric_eom_dbi_gtt})+$\f{g_{rr}}{g_{tt}}$ eq (\ref{detailed_metric_eom_dbi_dilaton_grr})=0. Using this one can show that ${\cal R}_{rr}+ \f{g_{rr}}{g_{tt}}{\cal R}_{tt}=0$. So it suggests that not all the equations of motion of the metric components are independent.  
Because of the relation between eq(\ref{detailed_metric_eom_dbi_gtt}) and eq (\ref{detailed_metric_eom_dbi_dilaton_grr}), let us assume that the metric component $g_{tt}$ and $g_{rr}$  are inversely related to each other. Moreover, we are interested to find AdS solution at the boundary suggests to consider the following form of the metric components 
\be
ds^2=\left(\f{r}{R}\right)^2\left[- f(r)dt ^2+dx^2_i\right]+\left(\f{R}{r}\right)^2  \f{dr^2}{f(r)}.
\ee
and the other fields as follows

\bea
\phi(r)&=&\phi_{0},\quad  V(\phi)=0,\quad W_1(\phi)=W_2(\phi)=1,\quad Z_1(\phi)=Z_2(\phi)=1
\eea
where $R$ is the AdS radius and $\phi_0$ is constant.

With such  a choice of the metric, let us calculate  eq(\ref{detailed_metric_eom_dbi_gtt})  explicitly 

\bea\label{2nd_order_dif_eq}
&&r~ f''(r)+6f'(r)+\f{2R^2\Lambda}{r}+\f{6f(r)}{r}+
T_b \f{r R^2}{\sqrt{r^4+R^4\rho^2}}=0.
\eea

and eq  (\ref{detailed_metric_eom_dbi_dilaton_gxx})
explicitly has the following form
\bea\label{1st_order_dif_eq}
r~f'(r)+3f(r)+\Lambda R^2+\f{k^2 R^4}{2r^2}+T_bR^2 \f{\sqrt{r^4+R^4\rho^2}}{2r^2}=0.
\eea

It is easy to notice that eq(\ref{2nd_order_dif_eq}) does not depend on the dissipation parameter, $k$, whereas eq(\ref{1st_order_dif_eq}) does. Moreover, one can  show that eq(\ref{2nd_order_dif_eq}) in fact follows from eq(\ref{1st_order_dif_eq}). The precise relation is as follows: $\f{1}{r} \f{\p}{\p r}\left( r^2 \times eq(\ref{1st_order_dif_eq})\right)=eq(\ref{2nd_order_dif_eq})$. It is easier to solve a first order differential equation eq(\ref{1st_order_dif_eq}) than a second order differential equation eq(\ref{2nd_order_dif_eq}). Hence, the solution is
\bea
f(r)&=&\f{c_1}{r^3}-\f{\Lambda R^2}{3}+\f{k^2 R^4}{2r^2}-T_b R^2\f{\sqrt{r^4+R^4\rho^2}}{6r^2}\nn&-&T_b \f{R^4\sqrt{\rho^2}}{3r^2}{}_2F_1\left[\f{1}{2}, \f{1}{4}, 1+\f{1}{4}, -\f{r^4}{R^4\rho^2} \right],
\eea
where $c_1$ is a constant and   ${}_2F_1[a,b,c,z]$ is the hypergeometric function.

\paragraph{Temperature, entropy density and chemical potential:}

\bea\label{temp}
T_H&=&-\f{r_h}{8\pi}\left[ 2\Lambda+\f{k^2R^2}{r^2_h}+\f{T_b}{r^2_h}\sqrt{r^4_h+R^4\rho^2}\right],\nn
\mu&=&-r_h ~{}_2F_1\Bigg[\f{1}{2},\f{1}{4},\f{5}{4},-\f{ r^{4}_h}{\rho^2R^{4}}\Bigg], ~s=\f{2\pi}{\kappa^2}\Bigg( \f{r_h}{R}\Bigg)^{2}.
\eea
 where $\mu_{DBI}$ is the chemical potential associated to the gauge potential.

\paragraph{Null energy condition:}
The null energy condition is given as $T_{MN}u^M u^N \geq 0$, where $ T_{MN} $ and $ u^{M} $ are the energy momentum tensor and the null vectors respectively. Using the Einstein's equation of motion for metric components, the null energy condition can be re-written as ${\cal R}_{MN}u^M u^N \geq 0$, where $ {\cal R}_{MN} $ is the Ricci tensor. 
The interesting choice\footnote{As the other choice $u^t=1/\sqrt{g_{tt}},~~ u^r=1/\sqrt{g_{tt}},~~u^{x_i}=0$ gives nothing interesting.} would be as  $u^t=1/\sqrt{g_{tt}},~~ u^r=0,~~u^{x_1}=1/\sqrt{g_{xx}}$ and $u^{x_2}=0$. In which case the null energy condition becomes 
 \bea
{\cal R}_{MN} u^M u^N&=&\f{r^2f''(r)+4r f'(r)}{2R^2}\nn
&=&\f{k^2R^{4}}{r^{2}}+ T_b R^6~ \f{\rho^2}{r^2\sqrt{r^4+\rho^2R^4}} \nn
&\geq & 0
\eea

For positive tension of the brane, $T_b>0$, the null energy condition is respected automatically.

\paragraph{ Solution at IR: }

At IR, it is expected that the metric typically exhibits the feature $g_{tt}\neq g_{rr}$. In which case, we shall write the geometry  as follows 
\be\label{ansatz_ir}
ds^2=-U_1(r)dt^2+\f{1}{U_2(r)}dr^2+h(r)(dx^2+dy^2).
\ee

Let us assume $W_1(\phi)=W_2(\phi)=\psi(\phi)$. 
The functions $U_1,~U_2$ and $h$  obeys the following differential equations 
\bea\label{diff_function}
&&2U_2 U''_1+U'_1 U'_2-\f{U_2 U'^2_1}{U_1}+\f{2U_2 h'U'_1}{h}+2U_1(2\Lambda+V)+2T_b \sqrt{U_1}Z_1Z_2\sqrt{U_1Z^2_2-U_2A'^2_t}=0,\nn
&&2 U_2 h''+h' U'_2+\f{U_2 h' U'_1}{U_1}+2k^2\psi+2h\left[2\Lambda+V+T_b\f{\sqrt{U_1}Z_1Z^3_2}{\sqrt{U_1Z^2_2-U_2A'^2_t}} \right]=0,\nn
&&2h^2U_1U_2 U''_1+4hU^2_1U_2 h''+2h^2U^2_1(2\Lambda+V+U_2\phi'^2)+U'_2hU_1(hU'_1+2U_1h')-\nn&&U_2(h^2U'^2_1+2U^2_1h'^2)+2T_b h^2 U^{3/2}_1Z_1Z_2 \sqrt{U_1Z^2_2-U_2A'^2_t}=0.
\eea

The equation of motion associated to scalar field, $\phi$ gives
\bea\label{scalar_diff_eq_ir}
&&h\phi''+\f{(2U_1U_2h'+hU_2U'_1+hU_1U'_2)}{2U_1U_2}\phi'-
\left(\f{h}{U_2}\f{dV}{d\phi}+\f{k^2}{U_2}\f{d\psi}{d\phi}\right)-\nn&&T_b \f{h}{\sqrt{U_1}U_2\sqrt{U_1Z^2_2-U_2 A'^2_t}}\left[U_1Z^2_2(Z_2Z'_1+2Z_1Z'_2)-U_2A'^2_t((Z_2Z'_1+Z_1Z'_2)) \right]=0.
\eea

The equation of motion associated to  gauge  potential gives
\be\label{gauge_field_ir}
 A'_t(r)=\f{\rho}{\sqrt{U_2}}
\f{Z_2\sqrt{U_1}}{\sqrt{Z^2_1Z^2_2h^{2}+\rho^2}}.
\ee

\subsection{Fluctuation and equations}

Let us fluctuate the background solution  in the  following way
\bea\label{fluctuation_geometry_without_b}
ds^2&=&-U_1(r)dt^2+\f{1}{U_2(r)}dr^2+h(r)(dx^2+dy^2)+2G_{tx}(t, r)dt dx+
2G_{rx}(r)dx dr\nn&+&2G_{ty}(t, r)dt dy +2G_{ry}(r)dr dy,\quad \chi_1=k~~x+\delta\chi_1(r), \quad \chi_2=k~~y+\delta\chi_2(r)\nn A&=&A_t(r)dt+A_x(t,r)dx+A_y(t,r)dy,
\eea
where the fluctuating part of the metric and other fields shall be considered infinitesimally. In particular, it has the following structure
\bea\label{flu_metric_gauge_withgout_b}
A_x(t,r)&=&-E_x t+a_x(r)+\xi_x ~t~ A_t(r),\quad A_y(t,r)=-E_y t+a_y(r)+\xi_y ~t~ A_t(r)\nn
G_{tx}(t,r)&=&h(r) h_{tx}(r)-t~ \xi_x~  U_1(r),\quad G_{ty}(t,r)=h(r) h_{ty}(r)-t~ \xi_y ~ U_1(r),\nn
G_{rx}&=&h(r) h_{rx}(r),\quad G_{ry}=h(r) h_{ry}(r),
\eea
where $\xi_i$'s  are related to  thermal gradients along the spatial directions.
With such  form of the geometry and matter fields, the equation of motion associated to the metric components  takes the following form

\bea\label{diff_pert_geometry}
&&4h U_2 h' h''_{tx}+ h'_{tx}\bigg[2h^2(V+2\Lambda)+2k^2h\psi+9U_2 h'^2+2hh'U'_2-h^2U_2\phi'^2+\nn&&2T_b\f{h^2Z_1Z^3_2\sqrt{U_1}}{\sqrt{U_1Z^2_2-U_2 A'^2_t}} \bigg]-  4k^2\psi h' h_{tx}+ 4T_b\f{U_2\sqrt{U_1}Z_1Z_2A'_t a'_x h'}{\sqrt{U_1Z^2_2-U_2 A'^2_t}} =0,\nn 
&&4h U_2 h' h''_{ty}+ h'_{ty}\bigg[2h^2(V+2\Lambda)+2k^2h\psi+9U_2 h'^2+2hh'U'_2-h^2U_2\phi'^2+\nn&&2T_b\f{h^2Z_1Z^3_2\sqrt{U_1}}{\sqrt{U_1Z^2_2-U_2 A'^2_t}} \bigg]-  4k^2\psi h' h_{ty}+ 4T_b\f{U_2\sqrt{U_1}Z_1Z_2A'_t a'_y h'}{\sqrt{U_1Z^2_2-U_2 A'^2_t}} =0,\nn
&&2E_x T_b \f{h \sqrt{U_1U_2} Z_1 Z^2_2 A'_t h'}{\sqrt{U_1Z^2_2-U_2 A'^2_t}}+2k^2 h U_1\sqrt{U_2}Z_2\psi h' h_{rx}-2kh U_1\sqrt{U_2}Z_2\psi h'\delta\chi'_1\nn&&-\xi_x\bigg[2T_b\f{hZ_1Z^2_2\sqrt{U_1}(hU_1Z^2_2+A_t U_2 A'_t h')}{\sqrt{U_2}\sqrt{U_1Z^2_2-U_2 A'^2_t}}+2k^2h \f{U_1}{\sqrt{U_2}} Z_2\psi +3U_1\sqrt{U_2}Z_2 h'^2\nn &&+
\f{h^2U_1Z_2(4\Lambda+2V-U_2\phi'^2)}{\sqrt{U_2}}\bigg]=0,\nn
&&2E_y T_b \f{h \sqrt{U_1U_2} Z_1 Z^2_2 A'_t h'}{\sqrt{U_1Z^2_2-U_2 A'^2_t}}+2k^2 h U_1\sqrt{U_2}Z_2\psi h' h_{ry}-2kh U_1\sqrt{U_2}Z_2\psi h'\delta\chi'_2\nn&&-\xi_y\bigg[2T_b\f{hZ_1Z^2_2\sqrt{U_1}(hU_1Z^2_2+A_t U_2 A'_t h')}{\sqrt{U_2}\sqrt{U_1Z^2_2-U_2 A'^2_t}}+2k^2h \f{U_1}{\sqrt{U_2}} Z_2\psi +3U_1\sqrt{U_2}Z_2 h'^2\nn &&+
\f{h^2U_1Z_2(4\Lambda+2V-U_2\phi'^2)}{\sqrt{U_2}}\bigg]=0.
\eea

 From eq(\ref{diff_function}), it follows that we can write the quantity
\bea
&&2h^2(V+2\Lambda)+2k^2 h\psi+9U_2h'^2+2hh'U'_2-h^2U_2\phi'^2+
2T_b\f{h^2Z_1Z^3_2\sqrt{U_1}}{\sqrt{U_1Z^2_2-U_2 A'^2_t}}\nn&=&8U_2h'^2+\f{2hh'}{U_1}\left(U'_2U_1-U_2U'_1 \right)
\eea

The first two equations of eq(\ref{diff_pert_geometry}) can be re-written using the above equation as

\bea\label{diff_htx_hty_with_rho}
&&2h U_2 h' h''_{tx}+ h'_{tx}\bigg[ 4U_2h'^2+\f{hh'}{U_1}\left(U'_2U_1-U_2U'_1 \right)\bigg]-  2k^2\psi h' h_{tx}+ 2T_b\f{U_2\sqrt{U_1}Z_1Z_2A'_t a'_x h'}{\sqrt{U_1Z^2_2-U_2 A'^2_t}} =0,\nn 
&&2h U_2 h' h''_{ty}+ h'_{ty}\bigg[4U_2h'^2+\f{hh'}{U_1}\left(U'_2U_1-U_2U'_1 \right) \bigg]-  2k^2\psi h' h_{ty}+ 2T_b\f{U_2\sqrt{U_1}Z_1Z_2A'_t a'_y h'}{\sqrt{U_1Z^2_2-U_2 A'^2_t}} =0.\nn
\eea
\subsection{Currents}

In this section, we shall present  the radially conserved electric as well as the heat currents required for the computation of the transport coefficients. We shall set $2\kappa^2=1$ for convenience.

\paragraph{Electric Currents:}

There exists two  radially conserved electric currents associated to the gauge potentials $A_i(t,r)$ for $i=x,~y$. The current that follows from the  gauge potentials $A_i(t,r)$ are as follows 
\bea
\label{electrical_current}
J^x(r)&=&-\f{T_b Z_1 Z_2 \sqrt{U_2}\left(h A'_t h_{tx}+U_1a'_x\right)}{\sqrt{ \left(U_1{Z_2}^2- U_2{A'_t}^2\right)}},\nn
J^y(r)&=&-\f{T_b Z_1 Z_2 \sqrt{U_2}\left(h A'_t h_{ty}+U_1a'_y\right)}{\sqrt{ \left(U_1{Z_2}^2- U_2{A'_t}^2\right)}}.
\eea

\paragraph{Heat Currents:}

The radially conserved heat currents are 
\bea\label{heat_current}
Q^x&=&U^{3/2}_1\sqrt{U_2}\p_r\left(\f{h h_{tx}}{U_1} \right)-A_t(r) J^x(r)=\sqrt{\f{U_2}{U_1}}\bigg( hU_1 h'_{tx}+h_{tx}\left[U_1h'-h U'_1\right]\bigg)-A_t(r) J^x(r),\nn
Q^y&=&U^{3/2}_1\sqrt{U_2}\p_r\left(\f{h h_{ty}}{U_1} \right)-A_t(r) J^y(r)=\sqrt{\f{U_2}{U_1}}\bigg( hU_1 h'_{ty}+h_{ty}\left[U_1h'-h U'_1\right]\bigg)-A_t(r) J^y(r).\nn
\eea

\subsubsection{Currents at the horizon}

Before evaluating the currents at the horizon, we need to know the corresponding in-falling boundary condition at the horizon. The in-falling boundary condition on different components of the gauge field as well as the metric components are 

\bea
a_x(r)&=&-\f{E_x}{U_0}~Log(r-r_h)+{\cal O}(r-r_h),\quad a_y(r)=-\f{E_y}{U_0}~Log(r-r_h)+{\cal O}(r-r_h),\nn
h_{tx}(r)&=& h_{tx}(r_h)-\xi_x \left(\f{U_1(r)}{h(r)~ U_0}\right)~Log(r-r_h)+{\cal O}(r-r_h), \nn 
h_{ty}(r)&=&h_{ty}(r_h)-\xi_y \left(\f{U_1(r)}{h(r)~ U_0}\right)~Log(r-r_h)+{\cal O}(r-r_h),
\eea
where $U_0$ is independent of $r$ and is defined in eq(\ref{temp_charge_density}). We demand that the function $U_1(r)$ and $U_2(r)$ at the horizon goes as follows:
\be\label{u1_u2_horizon}
 U_1(r)=U^{(0)}_1(r-r_h)+{\cal O}(r-r_h)^2,\quad  U_2(r)=U^{(0)}_2(r-r_h)+{\cal O}(r-r_h)^2.
 \ee

\paragraph{Hawking Temperature:} The temperature at IR can easily be computed and is given by
\be\label{temp_charge_density}
T_H=\f{1}{4\pi}\f{U'_1(r_h)\sqrt{U_2(r_h)}}{\sqrt{U_1(r_h)}}=\f{\sqrt{U^{(0)}_1U^{(0)}_2}}{4\pi}\equiv \f{U_0}{4\pi}.
\ee

\paragraph{ Metric components at the horizon:} The equation written down in eq(\ref{diff_htx_hty_with_rho}) shows there is no mixing of the metric components $h_{tx}$ with $h_{ty}$. This is due to the non-existence of magnetic field. At the horizon, using the in-falling boundary condition as written above gives rise to
\bea
h_{tx}(r_h)&=&-\left(\f{T_b \rho E_x h(r_h)\sqrt{U^{(0)}_1U^{(0)}_2}+(h(r_h))^2 U^{(0)}_1U^{(0)}_2\xi_x}{h(r_h)h(r_h)U_0 k^2\psi(\phi(r_h))}\right) \nn
h_{ty}(r_h)&=&-\left(\f{T_b \rho E_y h(r_h)\sqrt{U^{(0)}_1U^{(0)}_2}+(h(r_h))^2 U^{(0)}_1U^{(0)}_2\xi_y}{h(r_h)h(r_h)U_0 k^2\psi(\phi(r_h))}\right).
\eea

\paragraph{Currents at the horizon:} On imposition of  the in-falling boundary condition, the electric currents 
at the horizon takes the following form

\bea\label{electric_current_horizon}
J^x(r_h)&=&-T_b\rho h_{tx}(r_h)+\f{T_b E_x \sqrt{U^{(0)}_1U^{(0)}_2}\sqrt{\rho^2+(h(r_h)Z_1(\phi(r_h))Z_2(\phi(r_h)))^2}}{U_0 h(r_h)Z_2(\phi(r_h))} \nn
J^y(r_h)&=&-T_b\rho h_{ty}(r_h)+\f{T_b E_y \sqrt{U^{(0)}_1U^{(0)}_2}\sqrt{\rho^2+(h(r_h)Z_1(\phi(r_h))Z_2(\phi(r_h)))^2}}{U_0 h(r_h)Z_2(\phi(r_h))}
\eea

whereas the heat currents becomes

\bea\label{heat_current_horizon}
{\cal Q}^x(r_h)=-h(r_h) h_{tx}(r_h)\sqrt{U^{(0)}_1U^{(0)}_2}, \quad
{\cal Q}^y(r_h)=-h(r_h) h_{ty}(r_h)\sqrt{U^{(0)}_1U^{(0)}_2}
\eea

\subsection{Transport quantities}

The longitudinal electrical conductivity, thermo-electric and thermal conductivity for the system under study can be calculated  by considering the behavior of the metric components at the horizon. In which case, the transport quantities reads as follows 
\bea\label{gen_transport}
\sigma^{xx}(r_h)&=&\sigma^{yy}(r_h)=T_b\f{\sqrt{U^{(0)}_1U^{(0)}_2}\sqrt{\rho^2+(h(r_h)Z_1(\phi(r_h))Z_2(\phi(r_h)))^2}}{h(r_h)U_0 Z_2(\phi(r_h))}\nn&+&T^2_b\rho^2\f{\sqrt{U^{(0)}_1U^{(0)}_2}}{k^2h(r_h)U_0 \psi(\phi(r_h))}\nn
&=&T_b\f{\sqrt{\rho^2+(h(r_h)Z_1(\phi(r_h))Z_2(\phi(r_h)))^2}}{h(r_h) Z_2(\phi(r_h))}+\f{T^2_b\rho^2}{k^2h(r_h) \psi(\phi(r_h))},\nn
\alpha^{xx}(r_h)&=&\alpha^{yy}(r_h)=\left(\f{T_b \rho}{T_H}\right)\left(\f{U^{(0)}_1U^{(0)}_2}{k^2 U_0 \psi(\phi(r_h))}\right)= \f{4\pi T_b\rho}{k^2  \psi(\phi(r_h))},\nn
{\overline\alpha^{xx}(r_h)}&=&{\overline\alpha^{yy}(r_h)}=\left(\f{T_b \rho}{T_H}\right)\left(\f{U^{(0)}_1U^{(0)}_2}{k^2 U_0 \psi(\phi(r_h))}\right)= \f{4\pi T_b\rho}{k^2  \psi(\phi(r_h))},\nn
{\overline\kappa}^{xx}(r_h)&=&{\overline\kappa}^{yy}(r_h)=
\f{h(r_h)(U^{(0)}_1U^{(0)}_2)^{3/2}}{T_H k^2U_0\psi(\phi(r_h))}=16\pi^2 T_H\f{h(r_h)}{k^2 \psi(\phi(r_h))}.
\eea
A non-zero thermoelectric conductivity follows due to the interaction of the charge density with either the electric current or the heat current. 
The thermal conductivity at the horizon reads as
\bea\label{gen_kappa}
\kappa^{xx}(r_h)&=&\kappa^{yy}(r_h)={\overline\kappa}^{xx}-T_H\f{(\alpha^{xx})^2}{\sigma^{xx}}
=16\pi^2 T_H\f{h(r_h)}{k^2 \psi(\phi(r_h))}\nn&\times&\left(1-
\f{T_b\rho^2Z_2(\phi(r_h))}{T_b\rho^2Z_2(\phi(r_h))+k^2\psi(\phi(r_h))\sqrt{\rho^2+(h(r_h)Z_1(\phi(r_h))Z_2(\phi(r_h)))^2}} \right)\nn
&=& \f{16\pi^2 T_H h(r_h)\sqrt{\rho^2+(h(r_h)Z_1(\phi(r_h))Z_2(\phi(r_h)))^2}}{T_b\rho^2Z_2(\phi(r_h))+k^2\psi(\phi(r_h))\sqrt{\rho^2+(h(r_h)Z_1(\phi(r_h))Z_2(\phi(r_h)))^2}}.
\eea
The Seebeck coefficient is defined as the ratio of the electric field with the temperature gradient for vanishing electric current. For the present case
\be
S^{xx}=\f{\alpha^{xx}}{\sigma^{xx}}=\f{4\pi \rho h Z_2(\phi(r_h))}{k^2\psi(\phi(r_h))\sqrt{\rho^2+(h(r_h)Z_1(\phi(r_h))Z_2(\phi(r_h)))^2}+T_b\rho^2 Z_2(\phi(r_h))}.
\ee

The Peltier coefficient is defined as the ratio of the heat current with the 
electric current for vanishing thermal gradient. For the present case
\be
\Pi^{xx}=T_H\f{\alpha^{xx}}{\sigma^{xx}}=T_H S^{xx}.
\ee

It is easy to notice that all the transport quantities depend on four functions.  The functions like $\psi(\phi),~Z_1(\phi),~Z_2(\phi)$ and $h(r)$, apart from the charge density, $\rho$, the tension of the brane, $T_b$, and the temperature, $T_H$, of the black hole.

The transport quantities at UV for the present system  is calculated in \cite{Pal:2019bfw}.

\section{Examples}

We shall study two examples, where we shall be calculating the transport quantities. 
The examples that we shall study belongs to Lifshitz class of solution and the conformal to Lifshitz solution. The Lifshitz solutions are described by a parameter, $z$, which essentially describes the scaling of time, which can be  different from the spatial coordinates.

\subsection{Lifshitz Solution}
 With all the necessary equations, let us try to find an exact Lifshitz  solution. The geometry associated to Lifshitz solution admits the following form
\be
ds^2=-r^{2z}f(r)dt^2+r^2(dx^2+dy^2)+\f{dr^2}{r^2f(r)},
\ee

which means the functions used in the geometry takes the following form $U_1(r)=r^{2z}f(r), ~~U_2(r)=r^2 f(r)$ and $h(r)=r^2$. We assume the scalar field to behave logarithmically, $\phi(r)=\phi_0 ~~Log~r$, where $\phi_0$ is a constant. Substituting such form of the functions in eq(\ref{diff_function}) results in
\bea\label{diff_lifshitz}
&&r^2f''(r)+3(1+z)r f'(r)+2\Lambda+V+2z(z+2)f(r)+T_b\f{r^2 Z^2_1Z^3_2}{\sqrt{\rho^2+r^4Z^2_1Z^2_2}}=0,\nn
&&2r^3 f'(r)+r^2(V+2\Lambda)+2r^2(z+2)f(r)+k^2\psi+T_b Z_2\sqrt{\rho^2+r^4Z^2_1Z^2_2}=0,\nn
&&r^2f''(r)+3(1+z)r f'(r)+2\Lambda+V+(2z^2+4+\phi^2_0)f(r)+T_b\f{r^2 Z^2_1Z^3_2}{\sqrt{\rho^2+r^4Z^2_1Z^2_2}}=0.\nn
\eea

On comparing the first and last equation of eq(\ref{diff_lifshitz}) says $\phi^2_0=4(z-1)$. Let us impose a condition on potential energy associated to the scalar field, $\phi$, as $V(\phi)=0$ and $f(r)=1-\left(\f{r_h}{r} \right)^{z+2}$. In which case the consistent solution follows after setting $\psi(\phi)=\psi_0 e^{\f{\phi}{\sqrt{z-1}}}=\psi_0 r^2$. The algebraic equation involving the parameters like $\psi_0, T_b$ and $\Lambda$ are 
\bea
4+2 z+2\Lambda+T_b\sqrt{1+\rho^2}+k^2\psi_0=0,\quad 4z+2z^2+2\Lambda+\f{T_b}{\sqrt{1+\rho^2}}=0.
\eea

The solutions are
\be
\psi_0=\f{1}{k^2}\left[2z+2z^2-4-\f{T_b\rho^2}{\sqrt{1+\rho^2}}\right],\quad \Lambda=-\left[2z+z^2+\f{T_b}{2\sqrt{1+\rho^2}} \right].
\ee

The scalar field equation gives the same solution of $\psi_0$ as written above provided $\f{dZ_1}{d\phi}=-\f{2}{\sqrt{z-1}}r^{-4},~~\f{dZ_2}{d\phi}=\f{1}{\sqrt{z-1}}r^{2}$
and $\f{d\psi}{d\phi}=\f{\psi_0}{\sqrt{z-1}}r^{2}$. It means $\psi(\phi)=\psi_0 e^{\f{\phi(r)}{\sqrt{z-1}}}=\psi_0 r^2, ~~Z_1(\phi)=e^{-2\f{\phi(r)}{\sqrt{z-1}}}= r^{-4}$ and $Z_2(\phi)=e^{\f{\phi(r)}{\sqrt{z-1}}}=r^2$.

The tension of the brane need to set in such a way that $\psi_0$ should be positive, otherwise it will make the coefficient of the kinetic term associated to axion negative. 

The temperature associated to such a gravitational system is given by the Hawking tempretaure
\be
T_H=\f{(z+2)}{4\pi} r^{z}_h.
\ee

\paragraph{Transport quantites:} The temperature dependence of the transport quantities for the Lifshitz solution turns out to be
\bea\label{lifshitz_transport}
\sigma^{xx}(r_h)&=&\f{2T_b(z^2+z-2)(1+\rho^2)}{2(z^2+z-2)\sqrt{1+\rho^2}-T_b\rho^2} \left(\f{2+z}{4\pi}\right)^{\f{4}{z}}T^{-\f{4}{z}}_H,\nn
\alpha^{xx}(r_h)&=& \f{4\pi T_b \rho\sqrt{1+\rho^2}}{\left[2(z^2+z-2)\sqrt{1+\rho^2}-T_b\rho^2\right]}
\left(\f{4\pi T_H}{2+z}\right)^{-\f{2}{z}},\nn
\kappa^{xx}(r_h)&=&\f{8\pi^2 T_H}{(z^2+z-2)},\quad S^{xx}=\f{2\pi \rho}{(z^2+z-2)\sqrt{1+\rho^2}}\left(\f{4\pi T_H}{2+z}\right)^{\f{2}{z}},\nn
\Pi^{xx}&=&\f{2\pi \rho}{(z^2+z-2)\sqrt{1+\rho^2}}\left(\f{4\pi }{2+z}\right)^{\f{2}{z}}T^{\f{2+z}{z}}_H.
\eea
It is interesting to note that the translational symmetry breaking parameter, $k$, disappears completely from the transport quantities as written in eq(\ref{lifshitz_transport}). The electrical conductivity and thermo-electric conductivites are driven both by the charge density and temperature. Unlike electrical conductivity and thermo-electric conductivites the thermal conductivity  is driven only by the temperature.

\subsection{Hyperscaling violation Solution}

 This type of solutions are characterized by two parameters $z$, and $\theta\equiv2\gamma$. The parameter, $z$ says about the different in scaling behavior of the time coordinate with respect to the spatial coordinates,
whereas the second parameter  $\theta$ describes its conformal structure to the Lifshitz soution. It can also be interpreted as the parameter that violates the scaling symmetry.

The geometry is assumed to take the following form
\be
ds^2=r^{-2\gamma}\left(-r^{2z}f(r)dt^2+r^2[dx^2+dy^2]+\f{dr^2}{r^2f(r)} \right).
\ee

It means the functions that are defined in eq(\ref{ansatz_ir}) takes the following form
\be
U_1(r)=r^{2z-2\gamma}f(r),\quad U_2(r)=r^{2(1+\gamma)}f(r),\quad h(r)=r^{2(1-\gamma)}
\ee

We shall also assume the functions like $Z_1,~~Z_2,~~\psi$ and $V$ take the following form
\be
Z_1(\phi(r))=e^{-\alpha\phi(r)},\quad Z_2(\phi(r))=e^{\beta\phi(r)}, \quad \psi(\phi(r))=\psi_0 e^{\psi_1\phi(r)},\quad V(\phi(r))=m_1 e^{m_2\phi(r)},
\ee
where $\alpha,~~\beta,~~\psi_o,~~\psi_1,~~m_1$ and $m_2$ are constants. Interestingly, we can solve the equations if we adopt the scalar field to have the  logarithmic  behavior as
\be
\phi(r)=\phi_0~~ log~~r,
\ee

where $\phi_0$ is constant, in which case, the radial dependence of the functions are as follows
\be
Z_1=r^{-\alpha\phi_0},\quad Z_2=r^{\beta\phi_0}, \quad \psi=\psi_0 r^{\psi_1\phi_0},\quad V(\phi(r))=m_1 r^{m_2\phi_0}.
\ee

Let us assume that the function $f(r)=1-\left( \f{r_h}{r}\right)^{\eta}$. The equations eq(\ref{diff_function}), eq(\ref{scalar_diff_eq_ir}) and eq(\ref{gauge_field_ir}) can be  solved with the following choice of the constants
\bea
\Lambda&=&0,\quad \phi_0=2\sqrt{(1-\gamma)(z-1-\gamma)},\quad \eta=2+z-2\gamma,\quad \alpha=\f{(2-\gamma)}{\sqrt{(1-\gamma)(z-1-\gamma)}},\nn
 \psi_0&=&\f{1}{k^2}\left(2(z-1)(z+2-2\gamma)-\f{T_b\rho^2}{\sqrt{1+\rho^2}}\right),\quad m_1=-2(z-\gamma)(2+z-2\gamma)-\f{T_b}{\sqrt{1+\rho^2}},\nn
 \beta&=&\f{1}{\sqrt{(1-\gamma)(z-1-\gamma)}}= \psi_1,\quad m_2=\f{\gamma}{\sqrt{(1-\gamma)(z-1-\gamma)}}.
\eea

For these choices of the constants the radial dependence of the functions are 

\bea
f(r)&=&1-\left( \f{r_h}{r}\right)^{2+z-2\gamma},\quad \phi(r)=2\sqrt{(1-\gamma)(z-1-\gamma)}~log~r,\quad Z_1=r^{-2(2-\gamma)},\nn
Z_2&=&r^2,\quad V=-\left( 2(z-\gamma)(2+z-2\gamma)+\f{T_b}{\sqrt{1+\rho^2}}\right)r^{2\gamma},\nn
\psi&=&\f{1}{k^2}\left(2(z-1)(z+2-2\gamma)-\f{T_b\rho^2}{\sqrt{1+\rho^2}}\right) r^2,\quad A'_t=\f{\rho}{\sqrt{1+\rho^2}}r^{z-1-2\gamma}.
\eea

The temperature associated to such a gravitational system is given by the Hawking tempretaure
\be
T_H=\f{(z+2-2\gamma)}{4\pi} r^{z}_h.
\ee

\paragraph{Transport quantities:}The  temperature dependence of the transport quantities for such a solution takes the following form

\bea\label{hsv_transport_without_b}
\sigma^{xx}(r_h)&=&\f{2T_b(z-1)(z+2-2\gamma)(1+\rho^2)}{2(z-1)(z+2-2\gamma)\sqrt{1+\rho^2}-T_b\rho^2} \left(\f{2+z-2\gamma}{4\pi}\right)^{\f{4-2\gamma}{z}}T^{-\f{4-2\gamma}{z}}_H,\nn
\alpha^{xx}(r_h)&=& \f{4\pi T_b \rho\sqrt{1+\rho^2}}{\left(2(z-1)(z+2-2\gamma)\sqrt{1+\rho^2}-T_b\rho^2\right)}
\left(\f{4\pi T_H}{z+2-2\gamma}\right)^{-\f{2}{z}},\nn
\kappa^{xx}(r_h)&=&\f{8\pi^2 }{(z-1)(z+2-2\gamma)}\left(\f{4\pi}{z+2-2\gamma}\right)^{\f{-2\gamma}{z}}T^{\f{z-2\gamma}{z}}_H,\nn
S^{xx}&=&\f{\rho}{2(z-1)\sqrt{1+\rho^2}}\left(\f{4\pi}{z+2-2\gamma}\right)^{\f{z+2-2\gamma}{z}} T^{\f{2(1-\gamma)}{z}}_H,\nn
\Pi^{xx}&=&\f{\rho}{2(z-1)\sqrt{1+\rho^2}}\left(\f{4\pi}{z+2-2\gamma}\right)^{\f{z+2-2\gamma}{z}}T^{\f{z+2-2\gamma}{z}}_H.
\eea

Once again, we see that the electrical conductivity and thermo-electrical conductivies are driven by both the charge density as well as the temperature whereas the thermal conductivity is driven only by the temperature. All the transport quantities are completely independent of the dissipation parameter, $k$. Strange metal like behavior can be generated provided we set $z=4,~\gamma=0$.

\section{With magnetic field}

Now let us turn on magnetic field along with charge density, in which case,  the fieldstrength becomes $F=A_t'(r) dr\w dt+B dx\w dy$. Using the geometry as in eq(\ref{ansatz_ir}), the equation of motion that results in
\bea\label{diff_eq_b}
&&2U_2 U''_1+U'_1 U'_2-\f{U_2 U'^2_1}{U_1}+\f{2U_2 h'U'_1}{h}+2U_1(2\Lambda+V)+2T_b h \sqrt{U_1}Z_1Z^2_2\f{\sqrt{U_1Z^2_2-U_2A'^2_t}}{\sqrt{B^2+h^2Z^2_2}}=0,\nn
&&2 U_2 h''+h' U'_2+\f{U_2 h' U'_1}{U_1}+2k^2\psi+2h(2\Lambda+V)+2T_b\sqrt{U_1}Z_1Z^2_2\f{\sqrt{B^2+h^2Z^2_2}}{\sqrt{U_1Z^2_2-U_2A'^2_t}} =0,\nn
&&2h^2U_1U_2 U''_1+4hU^2_1U_2 h''+2h^2U^2_1(2\Lambda+V+U_2\phi'^2)+U'_2hU_1(hU'_1+2U_1h')-\nn&&U_2(h^2U'^2_1+2U^2_1h'^2)+2T_b h^3 U^{3/2}_1Z_1Z^2_2 \f{\sqrt{U_1Z^2_2-U_2A'^2_t}}{\sqrt{B^2+h^2Z^2_2}}=0.
\eea
The equation of motion associated to scalar field, $\phi$ gives
\bea\label{scalar_diff_eq_ir_b}
&&h\phi''+\f{(2U_1U_2h'+hU_2U'_1+hU_1U'_2)}{2U_1U_2}\phi'-
\left(\f{h}{U_2}\f{dV}{d\phi}+\f{k^2}{U_2}\f{d\psi}{d\phi}\right)-\nn&&T_b \f{\sqrt{B^2+h^2Z^2_2}}{\sqrt{U_1}U_2}\sqrt{U_1Z^2_2-U_2 A'^2_t}\f{dZ_1}{d\phi}+T_b\f{Z_1Z_2[-U_1(B^2+2h^2Z^2_2)+h^2U_2A'^2_t]}{\sqrt{U_1}U_2\sqrt{B^2+h^2Z^2_2}\sqrt{U_1Z^2_2-U_2 A'^2_t}}\f{dZ_2}{d\phi}=0.\nn
\eea

The equation of motion associated to  gauge  potential gives
\be\label{gauge_field_ir_b}
A'_t(r)=\f{\rho}{\sqrt{U_2}}
\f{Z_2\sqrt{U_1}}{\sqrt{Z^2_1(B^2+Z^2_2h^{2})+\rho^2}}.
\ee

\subsection{Fluctuation}

In the presence of magnetic field, let us fluctuate the background solution  in the  following way
\bea\label{fluctuation_geometry}
ds^2&=&-U_1(r)dt^2+\f{1}{U_2(r)}dr^2+h(r)(dx^2+dy^2)+2G_{tx}(t, r)dt dx+
2G_{rx}(r)dx dr\nn&+&2G_{ty}(t, r)dt dy +2G_{ry}(r)dr dy,\quad \chi_1=k~~x+\delta\chi_1(r), \quad \chi_2=k~~y+\delta\chi_2(r)\nn A&=&A_t(r)dt+\f{B}{2}\left(-ydx+xdy\right)+A_x(t,r)dx+A_y(t,r)dy,
\eea
where the fluctuating part of the metric and other fields shall be considered infinitesimally. In particular, it has the following structure
\bea\label{flu_metric_gauge}
A_x(t,r)&=&-E_x t+a_x(r)+\xi_x ~t~ A_t(r),\quad A_y(t,r)=-E_y t+a_y(r)+\xi_y ~t~ A_t(r)\nn
G_{tx}(t,r)&=&h(r) h_{tx}(r)-t~ \xi_x~  U_1(r),\quad G_{ty}(t,r)=h(r) h_{ty}(r)-t~ \xi_y ~ U_1(r),\nn
G_{rx}&=&h(r) h_{rx}(r),\quad G_{ry}=h(r) h_{ry}(r),
\eea
where $\xi_i$'s  are related to  thermal gradients along the spatial directions.
With such a form of the geometry and matter field, the equation of motion associated to the metric components  takes the following form

\bea\label{diff_pert_geometry_b}
&&h U_2  h''_{tx}-  h_{tx}\left[ k^2\psi +\f{T_bB^2Z_1Z^2_2\sqrt{U_1 }}{\sqrt{B^2+h^2Z^2_2}\sqrt{U_1Z^2_2-U_2 A'^2_t}}\right]+\nn&& h'_{tx}\bigg[\f{h^2(2V+4\Lambda-U_2\phi'^2)+2h(k^2\psi+h'U'_2)+9U_2 h'^2}{4h'}+T_b\f{hZ_1Z^2_2\sqrt{U_1}\sqrt{B^2+h^2Z^2_2}}{2h'\sqrt{U_1Z^2_2-U_2 A'^2_t}} \bigg]\nn&&+ T_b\f{\sqrt{U_1}Z_1Z^2_2}{\sqrt{B^2+h^2Z^2_2}\sqrt{U_1Z^2_2-U_2 A'^2_t}}\left[B(-E_y+\xi_yA_t)+hU_2A'_t(a'_x+Bh_{ry}) \right] =0,\nn 
&&h U_2  h''_{ty}-  h_{ty}\left[ k^2\psi +\f{T_bB^2Z_1Z^2_2\sqrt{U_1 }}{\sqrt{B^2+h^2Z^2_2}\sqrt{U_1Z^2_2-U_2 A'^2_t}}\right]+\nn&& h'_{ty}\bigg[\f{h^2(2V+4\Lambda-U_2\phi'^2)+2h(k^2\psi+h'U'_2)+9U_2 h'^2}{4h'}+T_b\f{hZ_1Z^2_2\sqrt{U_1}\sqrt{B^2+h^2Z^2_2}}{2h'\sqrt{U_1Z^2_2-U_2 A'^2_t}} \bigg]\nn&&- T_b\f{\sqrt{U_1}Z_1Z^2_2}{\sqrt{B^2+h^2Z^2_2}\sqrt{U_1Z^2_2-U_2 A'^2_t}}\left[B(-E_x+\xi_xA_t)-hU_2A'_t(a'_y-Bh_{rx}) \right] =0,\nn
&&k\psi\delta\chi'_1+\f{T_bZ_1Z^2_2[hA'_t(-E_x+Bh_{ty})+BU_1a'_y]}{\sqrt{U_1}\sqrt{B^2+h^2Z^2_2}\sqrt{U_1Z^2_2-U_2 A'^2_t}}-h_{rx}\left[k^2\psi+\f{T_bB^2Z_1Z^2_2\sqrt{U_1}}{\sqrt{B^2+h^2Z^2_2}\sqrt{U_1Z^2_2-U_2 A'^2_t}} \right]\nn&&+\xi_x\bigg[T_b\f{Z_1Z^2_2
	[U_1(B^2+h^2Z^2_2)+A_t hU_2 A'_t h']}{\sqrt{U_1}U_2h'\sqrt{B^2+h^2Z^2_2}\sqrt{U_1Z^2_2-U_2 A'^2_t}}+
\f{h^2(4\Lambda+2V-U_2\phi'^2)+2k^2h\psi+3U_2h'^2}{2hU_2h'}\bigg]=0,\nn
&&k\psi\delta\chi'_2-\f{T_bZ_1Z^2_2[hA'_t(E_y+Bh_{tx})+BU_1a'_x]}{\sqrt{U_1}\sqrt{B^2+h^2Z^2_2}\sqrt{U_1Z^2_2-U_2 A'^2_t}}-h_{ry}\left[k^2\psi+\f{T_bB^2Z_1Z^2_2\sqrt{U_1}}{\sqrt{B^2+h^2Z^2_2}\sqrt{U_1Z^2_2-U_2 A'^2_t}} \right]\nn&&+\xi_y\bigg[\f{T_bZ_1Z^2_2
	[U_1(B^2+h^2Z^2_2)+A_t hU_2 A'_t h']}{\sqrt{U_1}U_2h'\sqrt{B^2+h^2Z^2_2}\sqrt{U_1Z^2_2-U_2 A'^2_t}}+
\f{h^2(4\Lambda+2V-U_2\phi'^2)+2k^2h\psi+3U_2h'^2}{2hU_2h'}\bigg]=0.\nn
\eea

From eq(\ref{diff_eq_b}), it follows that we can write the quantity
\bea
&&\f{h^2(2V+4\Lambda-U_2\phi'^2)+2h(k^2\psi+h'U'_2)+9U_2 h'^2}{4h'}+T_b\f{hZ_1Z^2_2\sqrt{U_1}\sqrt{B^2+h^2Z^2_2}}{2h'\sqrt{U_1Z^2_2-U_2 A'^2_t}}\nn&=&2U_2h'+\f{h}{2U_1}\left(U'_2U_1-U_2U'_1 \right)
\eea

The first two equations of eq(\ref{diff_pert_geometry_b}) can be re-written using the above equation as

\bea\label{diff_htx_hty_b}
&&h U_2  h''_{tx}-  h_{tx}\left[ k^2\psi +\f{T_bB^2Z_1Z^2_2\sqrt{U_1 }}{\sqrt{B^2+h^2Z^2_2}\sqrt{U_1Z^2_2-U_2 A'^2_t}}\right]+ h'_{tx}\bigg[2U_2h'+\f{h}{2U_1}\left(U'_2U_1-U_2U'_1 \right) \bigg]\nn&&+ T_b\f{\sqrt{U_1}Z_1Z^2_2}{\sqrt{B^2+h^2Z^2_2}\sqrt{U_1Z^2_2-U_2 A'^2_t}}\left[B(-E_y+\xi_yA_t)+hU_2A'_t(a'_x+Bh_{ry}) \right] =0,\nn 
&&h U_2  h''_{ty}-  h_{ty}\left[ k^2\psi +\f{T_bB^2Z_1Z^2_2\sqrt{U_1 }}{\sqrt{B^2+h^2Z^2_2}\sqrt{U_1Z^2_2-U_2 A'^2_t}}\right]+ h'_{ty}\bigg[2U_2h'+\f{h}{2U_1}\left(U'_2U_1-U_2U'_1 \right) \bigg]\nn&&- T_b\f{\sqrt{U_1}Z_1Z^2_2}{\sqrt{B^2+h^2Z^2_2}\sqrt{U_1Z^2_2-U_2 A'^2_t}}\left[B(-E_x+\xi_xA_t)-hU_2A'_t(a'_y-Bh_{rx}) \right] =0.\nn
\eea

\subsection{Current}

The radially conserved electric current for the present case reads as   
\bea\label{electric_current_b}
J^x(r)&=&-\f{T_b hU_1\sqrt{U_2}Z_1Z^2_2(a'_x+Bh_{ry}) }{\sqrt{B^2+h^2Z^2_2}\sqrt{U_1Z^2_2-U_2 A'^2_t}}-\f{T_b \sqrt{U_2}Z_1A'_t[B(-E_y+\xi_y A_t)+h^2Z^2_2h_{tx}] }{\sqrt{B^2+h^2Z^2_2}\sqrt{U_1Z^2_2-U_2 A'^2_t}}\nn&-&BT_b\xi_y\int^r_{r_h}Z_1\f{\sqrt{U_1Z^2_2-U_2 A'^2_t}}{\sqrt{U_2}\sqrt{B^2+h^2Z^2_2}}
\nn
J^y(r)&=&-\f{T_b hU_1\sqrt{U_2}Z_1Z^2_2(a'_y-Bh_{rx}) }{\sqrt{B^2+h^2Z^2_2}\sqrt{U_1Z^2_2-U_2 A'^2_t}}+\f{T_b \sqrt{U_2}Z_1A'_t[B(-E_x+\xi_x A_t)-h^2Z^2_2h_{ty}] }{\sqrt{B^2+h^2Z^2_2}\sqrt{U_1Z^2_2-U_2 A'^2_t}}\nn&+&BT_b\xi_x\int^r_{r_h}Z_1\f{\sqrt{U_1Z^2_2-U_2 A'^2_t}}{\sqrt{U_2}\sqrt{B^2+h^2Z^2_2}}.
\eea

\paragraph{Heat currents:} Let us construct the radially conserved  heat currents associated to the geometric fluctuation. In order to do so, let us construct a quantity $Q^x$ as
\be
Q^x\equiv U_1^\f{3}{2}~  U_2^\f{1}{2}\p_r\left(\f{h~ h_{tx}}{U_1} \right)-A_t J^x.
\ee
The derivative of $Q^x$ using the conservation of the electric current, $\p_r J^x=0$,  gives
\be
\p_r Q^x=\p_r\left[U_1^\f{3}{2}~  U_2^\f{1}{2}\p_r\left(\f{h~ h_{tx}}{U_1} \right)\right]-A'_t J^x.
\ee
Using eq(\ref{diff_eq_b}) and eq(\ref{diff_pert_geometry_b}), the quantity, $\p_r Q^x$ becomes
\be
\p_r Q^x=\f{B T_b(E_y-\xi_y A_t) Z_1\sqrt{U_1Z^2_2-U_2 A'^2_t}}{\sqrt{U_2}\sqrt{\left(B^2 +h^2 {Z_2}^2\right) }}+\xi_y A'_t\int^r_{r_h}\f{B T_b Z_1\sqrt{U_1Z^2_2-U_2 A'^2_t}}{\sqrt{U_2}\sqrt{\left(B^2 +h^2 {Z_2}^2\right) }}\equiv {\cal M}^x(r).
\ee

It is easy to see that  the form of the conserved heat current along x-direction takes the following form
\be\label{heat_current_qx}
{\cal Q}^x(r)=U_1^\f{3}{2}~  U_2^\f{1}{2}\p_r\left(\f{h~ h_{tx}}{U_1} \right)-A_t J^x-\int^r_{r_h}dr' {\cal M}^x(r').
\ee

Similarly, defining $Q^y \equiv U_1^\f{3}{2}~  U_2^\f{1}{2}\p_r\left(\f{h~ h_{ty}}{U_1} \right)-A_t J^y$, one gets
\be
\p_r Q^y=-\f{B T_b(E_x-\xi_x A_t) Z_1\sqrt{U_1Z^2_2-U_2 A'^2_t}}{\sqrt{U_2}\sqrt{\left(B^2 +h^2 {Z_2}^2\right) }}-\xi_x A'_t\int^r_{r_h}\f{B T_b Z_1\sqrt{U_1Z^2_2-U_2 A'^2_t}}{\sqrt{U_2}\sqrt{\left(B^2 +h^2 {Z_2}^2\right) }}\equiv {\cal M}^y(r).
\ee

So the conserved heat current along y-direction, $\p_r{\cal Q}^y(r) =0$, is 
\be\label{heat_current_qy}
{\cal Q}^y(r)=U_1^\f{3}{2}~  U_2^\f{1}{2}\p_r\left(\f{h~ h_{ty}}{U_1} \right)-A_t J^y-\int^r_{r_h}dr' {\cal M}^y(r').
\ee

\subsection{At the horizon} 
We shall investigate the behavior of different fields and currents at the horizon, which in turn  allows us to compute the transport quantities at the horizon.
If we consider the background geometry to be a black hole then the function $U(r)$ has a zero at the horizon, $r_h$. Near to the horizon, it can be expanded as $U(r)=U_0 (r-r_h)+\cdots $, where $U_0$ is a constant, in which case, the Hawking temperature reads as $T_H=U_0/(4\pi)$.

\paragraph{In-falling boundary condition:} The in-falling boundary condition for different metric components and the gauge potentials at the horizon are as follows: 
\bea
a_x(r)&=&-\f{E_x}{U_0}~Log(r-r_h)+{\cal O}(r-r_h),\quad a_y(r)=-\f{E_y}{U_0}~Log(r-r_h)+{\cal O}(r-r_h),\nn
h_{tx}(r)&=& Uh_{rx}(r_h)-\xi_x \left(\f{U_1(r)}{h(r)~ U_0}\right)~Log(r-r_h)+{\cal O}(r-r_h), \nn 
h_{ty}(r)&=&Uh_{ry}(r_h)-\xi_y \left(\f{U_1(r)}{h(r)~ U_0}\right)~Log(r-r_h)+{\cal O}(r-r_h),
\eea
where $U(r)\equiv \sqrt{U_1(r)U_2(r)}=U_0 (r-r_h)+\cdots$. Note, $U_0$ is independent of $r$ and respects the relation  $U_0=\sqrt{U^{(0)}_1U^{(0)}_2}$.

\paragraph{Electric current at the horizon:} The electric current at the horizon reads as
\bea
J^x(r_h)&=&-\left(\f{T_b hU_1\sqrt{U_2}Z_1Z^2_2(a'_x+Bh_{ry}) }{\sqrt{B^2+h^2Z^2_2}\sqrt{U_1Z^2_2-U_2 A'^2_t}}\right)_{r_h}-\left(\f{T_b \sqrt{U_2}Z_1A'_t[B(-E_y+\xi_y A_t)+h^2Z^2_2h_{tx}] }{\sqrt{B^2+h^2Z^2_2}\sqrt{U_1Z^2_2-U_2 A'^2_t}}\right)_{r_h}
\nn
&=&\left(\f{T_b}{B^2+h^2Z^2_2}\left[\rho(B E_y-h^2Z^2_2h_{tx})+hZ_2(E_x-Bh_{ty})\sqrt{\rho^2+Z^2_1(B^2+h^2Z^2_2)} \right]\right)_{r_h}\nn 
J^y(r_h)&=&-\left(\f{T_b hU_1\sqrt{U_2}Z_1Z^2_2(a'_y-Bh_{rx}) }{\sqrt{B^2+h^2Z^2_2}\sqrt{U_1Z^2_2-U_2 A'^2_t}}\right)_{r_h}+\left(\f{T_b \sqrt{U_2}Z_1A'_t[B(-E_x+\xi_x A_t)-h^2Z^2_2h_{ty}] }{\sqrt{B^2+h^2Z^2_2}\sqrt{U_1Z^2_2-U_2 A'^2_t}}\right)_{r_h}\nn
&=&\left(\f{T_b}{B^2+h^2Z^2_2}\left[-\rho(B E_x+h^2Z^2_2h_{ty})+hZ_2(E_y+Bh_{tx})\sqrt{\rho^2+Z^2_1(B^2+h^2Z^2_2)} \right]\right)_{r_h},\nn 
\eea
where in the second equality we have invoked the in-falling  boundary condition at the horizon for different components of the metric and gauge fields.
\paragraph{Heat current at the horizon:} The heat current at the horizon reads as
\bea\label{heat_current_horizon_b}
{\cal Q}^x(r_h)=-h(r_h) h_{tx}(r_h)\sqrt{U^{(0)}_1U^{(0)}_2}, \quad
{\cal Q}^y(r_h)=-h(r_h) h_{ty}(r_h)\sqrt{U^{(0)}_1U^{(0)}_2}
\eea
where we have used eq(\ref{u1_u2_horizon}).

\paragraph{$h_{tx}$ and $h_{ty}$ at the horizon:} With the in-falling boundary condition, the components of the metric at the horizon obeys 
\bea
&&\xi_x U_0+h_{tx}\left(k^2\psi+T_b B^2Z_2\f{\sqrt{\rho^2+Z^2_1(B^2+h^2Z^2_2)}}{B^2+h^2Z^2_2}\right)-T_b \f{h B \rho Z^2_2}{B^2+h^2Z^2_2}h_{ty}+\nn &&T_bZ_2\f{\sqrt{\rho^2+Z^2_1(B^2+h^2Z^2_2)}}{B^2+h^2Z^2_2}\left( BE_y+\f{hE_x\rho Z_2}{\sqrt{\rho^2+Z^2_1(B^2+h^2Z^2_2)}}\right)=0\nn
&&\xi_y U_0+h_{ty}\left(k^2\psi+T_b B^2Z_2\f{\sqrt{\rho^2+Z^2_1(B^2+h^2Z^2_2)}}{B^2+h^2Z^2_2}\right)+T_b \f{h B \rho Z^2_2}{B^2+h^2Z^2_2}h_{tx}+\nn &&T_bZ_2\f{\sqrt{\rho^2+Z^2_1(B^2+h^2Z^2_2)}}{B^2+h^2Z^2_2}\left( -BE_x+\f{hE_y\rho Z_2}{\sqrt{\rho^2+Z^2_1(B^2+h^2Z^2_2)}}\right)=0. 
\eea  

The solution 
\bea\label{htx_hty_horizon_b}
h_{tx}(r_h)&=&\f{1}{k^4\psi^2(B^2+h^2Z^2_2)+T_bB^2Z_2[T_bZ_2(\rho^2+B^2Z^2_1)+2k^2\sqrt{\rho^2+Z^2_1(B^2+h^2Z^2_2)}]}\times\nn
&&\bigg(-\xi_xU_0[k^2\psi(B^2+h^2Z^2_2)+T_bB^2Z_2\sqrt{\rho^2+Z^2_1(B^2+h^2Z^2_2)}]-E_x T_b\rho hZ^2_2 \psi k^2\nn&&-BT_bU_0\rho hZ^2_2\xi_y-E_yBT_bZ_2[k^2\sqrt{\rho^2+Z^2_1(B^2+h^2Z^2_2)}+T_bZ_2(\rho^2+B^2Z^2_1)]\bigg)\nn
h_{ty}(r_h)&=&\f{1}{k^4\psi^2(B^2+h^2Z^2_2)+T_bB^2Z_2[T_bZ_2(\rho^2+B^2Z^2_1)+2k^2\sqrt{\rho^2+Z^2_1(B^2+h^2Z^2_2)}]}\times\nn
&&\bigg(-\xi_yU_0[k^2\psi(B^2+h^2Z^2_2)+T_bB^2Z_2\sqrt{\rho^2+Z^2_1(B^2+h^2Z^2_2)}]-E_y T_b\rho hZ^2_2 \psi k^2\nn&&+BT_bU_0\rho hZ^2_2\xi_x+E_xBT_bZ_2[k^2\sqrt{\rho^2+Z^2_1(B^2+h^2Z^2_2)}+T_bZ_2(\rho^2+B^2Z^2_1)]\bigg),\nn
\eea
where the right hand side of these two equations need to be evaluated at the horizon.
 
\subsection{Transport quantities}

Substituting the value of $h_{tx}$ and $h_{ty}$ at the horizon into the electric currents 
and heat currents gives
\bea\label{transport}
\sigma^{xx}(r_h)&=&\sigma^{yy}(r_h)\nn
&=&\left(\f{k^2T_bhZ_2\psi[T_bZ_2(\rho^2+B^2Z^2_1)+k^2\psi\sqrt{\rho^2+Z^2_1(B^2+h^2Z^2_2)}]}{k^4\psi^2(B^2+h^2Z^2_2)+T_bB^2Z_2[T_bZ_2(\rho^2+B^2Z^2_1)+2k^2\psi\sqrt{\rho^2+Z^2_1(B^2+h^2Z^2_2)}]}\right)_{r_h} \nn
\sigma^{xy}(r_h)&=&-\sigma^{yx}(r_h)\nn
&=&\left(\f{B\rho T_b[T^2_bZ^2_2(\rho^2+B^2Z^2_1)+2T_bk^2Z_2\psi\sqrt{\rho^2+Z^2_1(B^2+h^2Z^2_2)}+k^4\psi^2]}{k^4\psi^2(B^2+h^2Z^2_2)+T_bB^2Z_2[T_bZ_2(\rho^2+B^2Z^2_1)+2k^2\psi\sqrt{\rho^2+Z^2_1(B^2+h^2Z^2_2)}]}\right)_{r_h}\nn
T_H \alpha^{xx}(r_h)&=&T_H \alpha^{yy}(r_h)\nn
&=&\left(\f{k^2T_bU_0\rho h^2Z^2_2\psi}{k^4\psi^2(B^2+h^2Z^2_2)+T_bB^2Z_2[T_bZ_2(\rho^2+B^2Z^2_1)+2k^2\psi\sqrt{\rho^2+Z^2_1(B^2+h^2Z^2_2)}]}\right)_{r_h}\nn
T_H \alpha^{xy}(r_h)&=&-T_H \alpha^{yx}(r_h)\nn
&=&\left(\f{B T_bhU_0Z_2[T_bZ_2(\rho^2+B^2Z^2_1)+k^2\psi\sqrt{\rho^2+Z^2_1(B^2+h^2Z^2_2)}]}{k^4\psi^2(B^2+h^2Z^2_2)+T_bB^2Z_2[T_bZ_2(\rho^2+B^2Z^2_1)+2k^2\psi\sqrt{\rho^2+Z^2_1(B^2+h^2Z^2_2)}]}\right)_{r_h}\nn
T_H{\overline\kappa}^{xx}(r_h)&=&T_H{\overline\kappa}^{yy}(r_h)\nn
&=&\left(\f{U^2_0h[k^2\psi(B^2+h^2Z^2_2)+B^2T_bZ_2\sqrt{\rho^2+Z^2_1(B^2+h^2Z^2_2)}]}{k^4\psi^2(B^2+h^2Z^2_2)+T_bB^2Z_2[T_bZ_2(\rho^2+B^2Z^2_1)+2k^2\psi\sqrt{\rho^2+Z^2_1(B^2+h^2Z^2_2)}]}\right)_{r_h} \nn
T_H{\overline\kappa}^{xy}(r_h)&=&-T_H{\overline\kappa}^{yx}(r_h)\nn
&=&\left(\f{\rho BT_bU^2_0h^2Z^2_2}{k^4\psi^2(B^2+h^2Z^2_2)+T_bB^2Z_2[T_bZ_2(\rho^2+B^2Z^2_1)+2k^2\psi\sqrt{\rho^2+Z^2_1(B^2+h^2Z^2_2)}]}\right)_{r_h} \nn
\eea
This gives the longitudinal thermal conductivity as
\bea\label{kappa_xx}
\kappa_{xx}(r_h)&=&\kappa_{yy}(r_h)=\Biggl[{\overline\kappa}_{xx}-T_H\f{\left((\alpha^2_{xx}-\alpha^2_{xy})\sigma_{xx}+2\alpha_{xx}\alpha_{xy}\sigma_{xy}\right)}{\sigma^2_{xx}+\sigma^2_{xy}}\Biggr]_{r_h}\nn
&=&\f{16\pi^2 T_H h\left[T_b \rho^2 Z_2\sqrt{\rho^2+Z^2_1(B^2+h^2 Z^2_2)}+k^2\psi(\rho^2+h^2Z^2_1Z^2_2)\right]}{ [2k^2 T_b\rho^2 Z_2\psi\sqrt{\rho^2+Z^2_1(B^2+h^2 Z^2_2)}+T^2_b\rho^2Z^2_2 (\rho^2+B^2 Z^2_1)+k^4\psi^2(\rho^2+h^2Z^2_1Z^2_2)]}\nn
\eea
The right hand side need to be evaluated at the horizon. The  transverse thermal conductivity takes the following form

\bea\label{kappa_xy}
\kappa_{xy}(r_h)&=&-\kappa_{yx}(r_h)=\Biggl[{\overline\kappa}_{xy}+T_H\f{\left((\alpha^2_{xx}-\alpha^2_{xy})\sigma_{xy}-2\alpha_{xx}\alpha_{xy}\sigma_{xx}\right)}{\sigma^2_{xx}+\sigma^2_{xy}}\Biggr]_{r_h}\nn&=&-\f{16 \pi^2 T_H B T_b \rho h^2Z^2_1Z^2_2}{ 2k^2 T_b\rho^2 Z_2\psi\sqrt{\rho^2+Z^2_1(B^2+h^2 Z^2_2)}+T^2_b\rho^2Z^2_2 (\rho^2+B^2 Z^2_1)+k^4\psi^2(\rho^2+h^2Z^2_1Z^2_2)}.\nn
\eea
 It is interesting to note that the transport quantities are independent of the form of potential energy, $V(\phi)$. However, for functions $U_1(r)$ and $U_2(r)$, we need only its derivative  at the horizon.

\subsection{An example: Hyperscaling violation Solution}

This type of solutions are characterized by two parameters $z$, and $\theta\equiv2\gamma$. The parameter, $z$ says about the different in scaling behavior of the time coordinate with respect to the spatial coordinates,
whereas the second parameter  $\theta$ describes it conformal structure to the Lifshitz soution. It can also be interpreted as the parameter that violates the scaling symmetry \cite{Huijse:2011ef}.

The geometry is assumed to take the following form
\be
ds^2=r^{-2\gamma}\left(-r^{2z}f(r)dt^2+r^2[dx^2+dy^2]+\f{dr^2}{r^2f(r)} \right).
\ee

It means the functions that are defined in eq(\ref{ansatz_ir}) takes the following form
\be
U_1(r)=r^{2z-2\gamma}f(r),\quad U_2(r)=r^{2(1+\gamma)}f(r),\quad h(r)=r^{2(1-\gamma)}
\ee

We shall also assume the functions like $Z_1,~~Z_2,~~\psi$ and $V$ take the following form
\be
Z_1(\phi(r))=e^{-\alpha\phi(r)},\quad Z_2(\phi(r))=e^{\beta\phi(r)}, \quad \psi(\phi(r))=\psi_0 e^{\psi_1\phi(r)},\quad V(\phi(r))=m_1 e^{m_2\phi(r)},
\ee
where $\alpha,~~\beta,~~\psi_0,~~\psi_1,~~m_1$ and $m_2$ are constants. Interestingly, we can solve the equations if we adopt the scalar field to have the  logarithmic  behavior as
\be
\phi(r)=\phi_0~~ log~~r,
\ee

where $\phi_0$ is constant. In which case the radial dependence of the functions are as follows
\be
Z_1=r^{-\alpha\phi_0},\quad Z_2=r^{\beta\phi_0}, \quad \psi=\psi_0 r^{\psi_1\phi_0},\quad V(\phi(r))=m_1 r^{m_2\phi_0}.
\ee

Let us assume that the function $f(r)=1-\left( \f{r_h}{r}\right)^{\eta}$. Equations eq(\ref{diff_eq_b}), eq(\ref{scalar_diff_eq_ir_b}) and eq(\ref{gauge_field_ir_b}) can be  solved with the following choice of the constants
\bea
\Lambda&=&0,\quad\gamma=2,\quad \phi_0=2\sqrt{(3-z)},\quad \eta=z-2,\quad \alpha=0,\nn
\psi_0&=&\f{1}{k^2}\left(2(z-1)(z-2)-\f{T_b(\rho^2+B^2)}{\sqrt{1+\rho^2+B^2}}\right),\quad m_1=-2(z-2)^2-\f{T_b}{\sqrt{1+\rho^2+B^2}},\nn
\beta&=&\f{1}{\sqrt{3-z}}= \psi_1,\quad m_2=\f{2}{\sqrt{3-z}}.
\eea

For these choices of the constants the radial dependence of the functions are 

\bea
f(r)&=&1-\left( \f{r_h}{r}\right)^{z-2},\quad \phi(r)=2\sqrt{(3-z)}~log~r,\quad Z_1=1,\quad
Z_2=r^2,\nn V&=&-\left( 2(z-2)^2+\f{T_b}{\sqrt{1+\rho^2+B^2}}\right)r^{4},\nn
\psi&=&\f{1}{k^2}\left(2(z-1)(z-2)-\f{T_b(\rho^2+B^2)}{\sqrt{1+\rho^2+B^2}}\right) r^2,\quad A'_t=\f{\rho}{\sqrt{1+\rho^2+B^2}}r^{z-3}.
\eea

The temperature associated to such a gravitational system is given by the Hawking tempretaure
\be
T_H=\f{(z-2)}{4\pi} r^{z}_h.
\ee

\paragraph{Transport quantities:}The  temperature dependence of the transport quantities for such a solution takes the following form

\bea\label{hsv_transport}
\sigma^{xx}(r_h)&=&\f{T_b k^2\psi_0\left(T_b(\rho^2+B^2)+\psi_0k^2\sqrt{1+\rho^2+B^2}\right)}{T^2_bB^2(\rho^2+B^2)+k^4\psi^2_0(1+B^2)+2T_bk^2\psi_0B^2\sqrt{1+\rho^2+B^2}},\nn
\sigma^{xy}(r_h)&=&\f{T_b\rho B \left(T^2_b(\rho^2+B^2)+2\psi_0k^2T_b\sqrt{1+\rho^2+B^2}+k^4\psi^2_0\right)}{T^2_bB^2(\rho^2+B^2)+k^4\psi^2_0(1+B^2)+2T_bk^2\psi_0B^2\sqrt{1+\rho^2+B^2}},\nn
\alpha^{xx}(r_h)&=& \f{4\pi\rho k^2\psi_0 T_b}{T^2_bB^2(\rho^2+B^2)+k^4\psi^2_0(1+B^2)+2T_bk^2\psi_0B^2\sqrt{1+\rho^2+B^2}}
\left(\f{4\pi T_H}{z-2}\right)^{-\f{2}{z}},\nn
\alpha^{xy}(r_h)&=& \f{4\pi B T_b\left(T_b(\rho^2+B^2)+k^2\psi_0\sqrt{1+\rho^2+B^2}\right)}{T^2_bB^2(\rho^2+B^2)+k^4\psi^2_0(1+B^2)+2T_bk^2\psi_0B^2\sqrt{1+\rho^2+B^2}}
\left(\f{4\pi T_H}{z-2}\right)^{-\f{2}{z}},\nn
{\overline\kappa}^{xx}(r_h)&=&\f{16\pi^2 \left(k^2\psi_0(1+B^2)+T_bB^2\sqrt{1+\rho^2+B^2}\right)}{T^2_bB^2(\rho^2+B^2)+k^4\psi^2_0(1+B^2)+2T_bk^2\psi_0B^2\sqrt{1+\rho^2+B^2}}
\left(\f{4\pi }{z-2}\right)^{-\f{4}{z}}T^{{\f{z-4}{z}}}_H,\nn
{\overline\kappa}^{xy}(r_h)&=&\f{16\pi^2 \rho BT_b}{T^2_bB^2(\rho^2+B^2)+k^4\psi^2_0(1+B^2)+2T_bk^2\psi_0B^2\sqrt{1+\rho^2+B^2}}
\left(\f{4\pi }{z-2}\right)^{-\f{4}{z}}T^{{\f{z-4}{z}}}_H\nn.
\eea

Once again, we see that the  thermo-electrical conductivities and thermal conductivities are driven by the charge density, magnetic field as well as the temperature whereas the electrical conductivities are independent of temperature.  All the transport quantities are completely independent of the dissipation parameter, $k$.

\subsection{A cooked-up model}

In the presence of magnetic field the electrical conductivity did not show  any temperature dependence as studied  in the   previous section makes it very hard to draw any useful conclusion. In this subsection, we shall cooked-up a model for the DBI action where the geometry is  assumed to take the following form
\be
ds^2=r^{-2\gamma}\left(-r^{2z}f(r)dt^2+r^2[dx^2+dy^2]+\f{dr^2}{r^2f(r)} \right).
\ee

It means the functions that are defined in eq(\ref{ansatz_ir}) takes the following form
\be
U_1(r)=r^{2z-2\gamma}f(r),\quad U_2(r)=r^{2(1+\gamma)}f(r),\quad h(r)=r^{2(1-\gamma)}
\ee

We shall also assume the functions like $Z_1,~~Z_2,$ and $\psi$  take the following form
\be
Z_1(\phi(r))=z_1 e^{-\alpha\phi(r)},\quad Z_2(\phi(r))=z_2 e^{\beta\phi(r)}, \quad \psi(\phi(r))=\psi_0 {\tilde\psi}_0 e^{\psi_1\phi(r)},\quad 
\ee
where $z_1,~~z_2,~~\alpha,~~\beta,~~\psi_0,~~{\tilde\psi}_0$ and $~~\psi_1$  are constants. Some of the  constants shall be fixed up by having the desired temperature dependence of the transport quantities. Let us consider the logarithmic behavior of the scalar field: $\phi(r)=\phi_0~~ log~~r$,
where $\phi_0$ is constant, in which case, the radial dependence of the functions are as follows
\be
Z_1=z_1 r^{-\alpha\phi_0},\quad Z_2=z_2 r^{\beta\phi_0}, \quad \psi=\psi_0 {\tilde\psi}_0 r^{\psi_1\phi_0}.
\ee  

The temperature is related to the horizon via dimensional analysis as \cite{Hartnoll:2009ns}
\be
r_h=A~ T^{1/z}_H,\quad {\rm where}\quad {\rm A\quad is \quad a\quad constant}
\ee

It means 
\bea
Z_1(\phi(r_h))&=&z_1 r^{-\alpha\phi_0}_h=T^{-\alpha\phi_0/z}_H,\quad Z_2(\phi(r_h))=z_2 r^{\beta\phi_0}_h=T^{\beta\phi_0/z}_H, \nn \psi(\phi(r_h))&=&\psi_0 {\tilde\psi}_0 r^{\psi_1\phi_0}_h=\psi_0T^{\psi_1\phi_0/z}_H,\nn h(r_h)&=& r^{2(1-\gamma)}_h=A^{\f{2(1-\gamma)}{z}}T^{\f{2(1-\gamma)}{z}}_H\equiv{\tilde A}T^{\f{2(1-\gamma)}{z}}_H
\eea

where we have set $z_1=A^{\alpha\phi_0},~z_2=A^{\beta\phi_0},~{\tilde\psi}_0=A^{-\psi_1\phi_0}$. The transport quantities reads as
\bea
\sigma^{xx}&=&{\tilde A}T_b \psi_0 k^2 T^{\f{2-2\gamma+\phi_0\psi_1}{z}}_H\f{\sigma_1}{\sigma_2},\quad {\rm where}\nn
\sigma_1&=&\left(T_bB^2+T_b\rho^2T^{\f{2\alpha\phi_0}{z}}_H +k^2 \psi_0T^{\f{(2\alpha-\beta+\psi_1)\phi_0}{z}}_H\sqrt{\rho^2+B^2T^{-\f{2\alpha\phi_0}{z}}_H+{\tilde A}^2T^{\f{4-4\gamma-2\alpha\phi_0+2\beta\phi_0}{z}}_H}\right),\nn
	\sigma_2&=&T^2_bB^4+{\tilde A}^2\psi^2_0k^4T^{\f{4-4\gamma+2(\alpha+\psi_1)\phi_0}{z}}_H\nn&+&B^2T^\f{2\alpha\phi_0}{z}_H\left(T^2_b\rho^2+\psi^2_0k^4T^{\f{2(\psi_1-\beta)\phi_0}{z}}_H+2T_b\psi_0k^2T^{\f{(\psi_1-\beta)\phi_0}{z}}_H\sqrt{\rho^2+B^2T^{-\f{2\alpha\phi_0}{z}}_H+{\tilde A}^2T^{\f{4-4\gamma-2\alpha\phi_0+2\beta\phi_0}{z}}_H}\right),\nn
\sigma^{xy}&=&T_b B\rho\f{\sigma_3}{\sigma_2},\quad {\rm where}\nn
\sigma_3&=&T^2_bB^2+T^2_b\rho^2T^{\f{2\alpha\phi_0}{z}}_H +2k^2 \psi_0T_bT^{\f{(2\alpha-\beta+\psi_1)\phi_0}{z}}_H\sqrt{\rho^2+B^2T^{-\f{2\alpha\phi_0}{z}}_H+{\tilde A}^2T^{\f{4-4\gamma-2\alpha\phi_0+2\beta\phi_0}{z}}_H},\nn
&+&\psi^2_0k^4 T^{\f{2(\alpha-\beta+\psi_1)\phi_0}{z}}_H,\nn
\alpha^{xx}&=&\f{4\pi T_b\rho\psi_0k^2 {\tilde A}^2}{\sigma_2}T^{\f{4-4\gamma+2\alpha\phi_0+\psi_1\phi_0}{z}}_H,\nn
\alpha^{xy}&=&4\pi B {\tilde A} T_b T^{\f{2-2\gamma}{z}}_H\f{\sigma_1}{\sigma_2},\quad \kappa^{xx}=\f{\kappa_1}{\kappa_2},\quad \kappa^{xy}=-\f{16\pi^2 B\rho{\tilde A}^2T_b}{\kappa_2}T^{\f{4+z-4\gamma}{z}}_H,\nn
\kappa_1&=&16\pi^2 {\tilde A} T^{\f{2+z-2\gamma}{z}}_H\bigg(T_b \rho^2T^{\f{(2\alpha-\beta)\phi_0}{z}}_H \sqrt{\rho^2+B^2T^{-\f{2\alpha\phi_0}{z}}_H+{\tilde A}^2T^{\f{4-4\gamma-2\alpha\phi_0+2\beta\phi_0}{z}}_H}+\psi_0k^2{\tilde A}^2T^{\f{4-4\gamma+\psi_1\phi_0}{z}}_H\nn&+&\psi_0k^2\rho^2T^{\f{(2\alpha-2\beta+\psi_1)\phi_0}{z}}_H\bigg),\nn
\kappa_2&=& B^2T^2_b\rho^2+T^2_b\rho^4T^{\f{2\alpha\phi_0}{z}}_H+2\psi_0k^2T_b \rho^2T^{\f{(2\alpha-\beta+\psi_1)\phi_0}{z}}_H \sqrt{\rho^2+B^2T^{-\f{2\alpha\phi_0}{z}}_H+{\tilde A}^2T^{\f{4-4\gamma-2\alpha\phi_0+2\beta\phi_0}{z}}_H}\nn&+&\psi^2_0k^4\left({\tilde A}^2T^{\f{2(2-2\gamma+\psi_1\phi_0)}{z}}_H+\rho^2T^{\f{2(\alpha-\beta+\psi_1)\phi_0}{z}}_H\right),\nn
\eea

Substituting these expressions in eq(\ref{transport}), eq(\ref{kappa_xx}) and eq(\ref{kappa_xy}) yields the temperature dependence of the transport quantities. Moreover as it stands  the transport quantities has a very complicated form and is very difficult to draw any precise conclusion. So, we shall find the temperature dependence of the transport quantities in the limit of small magnetic field and low density. To leading order, it takes the following form
\bea
\sigma^{xx}&=&T_b T^{-\f{\alpha\phi_0}{z}}_H+{\cal O}(\rho^2,B^2),\quad \sigma^{xy}=\f{2T^2_b}{{\tilde A}\psi_0k^2} \rho BT^{-\f{2(1-\gamma)+(\alpha+\psi_1)\phi_0}{z}}_H+{\cal O}(\rho^2,B^2),\nn \alpha^{xx}&=&\f{4\pi T_b}{\psi_0k^2}\rho T^{-\f{\psi_1\phi_0}{z}}_H+{\cal O}(\rho^2,B^2),\quad \alpha^{xy}=\f{4\pi T_b}{\psi_0k^2}B T^{-\f{(\alpha+\psi_1)\phi_0}{z}}_H+{\cal O}(\rho^2,B^2),\nn\kappa^{xx}&=&\f{16\pi^2{\tilde A}}{\psi_0k^2} T^{\f{2+z-2\gamma-\psi_1\phi_0}{z}}_H+{\cal O}(\rho^2,B^2),\quad\kappa^{xy}=-\f{16\pi^2T_b}{\psi^2_0k^4}\rho B T^{1-\f{2\psi_1\phi_0}{z}}_H+{\cal O}(\rho^2,B^2),\nn
\eea

It is interesting to note that the transport quantities depends only on five constants: $\alpha,~\psi_1,~\phi_0,~z$ and $\gamma$ to leading order in magnetic field and charge density. Now, we shall consider different cases, where we shall fix the constants in such a way  so as to read out the temperature dependence of the transport quantities which can resemble that of the strange metal.

\paragraph{Case I:} Let us fix the constants as 

\be\label{put_by_hand_I}
\alpha\phi_0=2(2-\gamma),\quad \psi_1\phi_0=2,\quad \gamma=0,\quad z=4
\ee

This gives the following temperature dependence of the transport quantities:

\bea
\sigma^{xx}&\sim& T^{-1}_H,\quad \sigma^{xy}\sim \rho BT^{-2}_H,\quad\kappa^{xx}\sim T_H,\quad\kappa^{xy}\sim-\rho B T^{0}_H,\nn \alpha^{xx}&\sim&\rho T^{-\f{1}{2}}_H,\quad \alpha^{xy}\sim B T^{-3/2}_H,\nn
\eea

This result is consistent with the result obtained using scaling analysis in section 2. Since, the precise temperature dependence of the transverse transport quantities are  not reproduced  allows us to make some further fixing of the constants. 

\paragraph{Case II:} Upon demanding the following conditions on the constants like $\alpha,~\psi_1,~\phi_0$ and $\gamma$, which we have put by hand, as
\be\label{put_by_hand_II}
\alpha\phi_0=z=\psi_1\phi_0,\quad 2-2\gamma=z
\ee
gives the temperature dependence of the transport quantities as

\bea
\sigma^{xx}&\sim& T^{-1}_H,\quad \sigma^{xy}\sim \rho BT^{-3}_H,\quad\kappa^{xx}\sim T_H,\quad\kappa^{xy}\sim-\rho B T^{-1}_H,\nn \alpha^{xx}&\sim&\rho T^{-1}_H,\quad \alpha^{xy}\sim B T^{-2}_H,\nn
\eea

  Eq(\ref{put_by_hand_II}) is constructed in such a way so as to have the precise  temperature dependence of electrical and thermal conductivity as that is required for strange metal. However, the temperature dependence of thermo-electric conductivity did not agree precisely. Let us mention {\it en passant} that it is not possible to fix all the constants so that  we can have the precise temperature dependence of all the transport quantities as required for the strange metal. In the present case, at least, we could manage to show the precise temperature dependence of four  transport quantities.

By going through the result both without the magnetic field and with the magnetic field, it follows that in order to have the desired temperature dependence of the transport quantities as in the case of strange metal, we need to consider   the dilaton dependent axionic  kinetic energy term must be proportional to the entropy density, $\psi(\phi(r_h))\sim h(r_h)\sim s$, where $s$ is the entropy density.

\paragraph{Null Energy Condition:} 
The covariant definition of null energy condition is $R_{MN}u^Mu^N\geq 0$ for some null vector $u^M$. By considering the null vector as $u^M=\left(\f{r^{2(\gamma-z)}}{\sqrt{f(r)}},0,0,r^{2(1+\gamma)}\sqrt{f(r)}\right)$, gives the condition  as  $(\gamma-1)(1-z+\gamma)f(r)\geq 0$. For the two choices as written above as case I and case II satisfies the null energy condition provided $f(r)\geq 0,$ which is usually 
the case, if one tries to connect the zero temperature solution with non-zero temperature solution.

\begin{table}
	\begin{center}
		\begin{tabular}{ |l |p{1.5cm}|p{1.5cm}| p{1.5cm} | p{1.5cm}| p{2.5cm} |}
			\hline
			Transport quantities in  &without&Scaling&with & with  & Strange metal \\ 
			Einstein-DBI system &magnetic field,&analysis&magnetic field,&magnetic field,&(as observed)\\
			&$B = 0$& eq(\ref{scaling_transport}) &$B \neq 0$	&$B \neq 0$&\\
		 &(Exact)&suggests& (case I)&  (case II)&\\\hline\hline
			Longitudinal electrical &&&&&\\ conductivity, $\sigma_{xx}=\sigma_{yy}$ &$ T^{-1}_H$&$ T^{-1}_H$&$ T^{-1}_H$&$ T^{-1}_H$& $ T^{-1}_H$\\\hline
			Transverse electrical &&&&&\\ conductivity, $\sigma_{xy}=-\sigma_{yx}$&&$ \rho BT^{-2}_H$ &$ \rho BT^{-2}_H$&$ \rho BT^{-3}_H$& $ T^{-3}_H$\\\hline
			Longitudinal thermoelectric &&&&&\\ conductivity, $\alpha_{xx}=\alpha_{yy}$ &$ \rho T^{-\f{1}{2}}_H$&$ \rho T^{-\f{1}{2}}_H$&$ \rho T^{-\f{1}{2}}_H$&$ \rho T^{-1}_H$&$ \rho T^{-\f{1}{2}}_H$\\\hline
			Transverse thermoelectric &&&&&\\ conductivity, $\alpha_{xy}=-\alpha_{yx}$&&$  BT^{-\f{3}{2}}_H$ &$  BT^{-\f{3}{2}}_H$&$  BT^{-2}_H$&$ B T^{-\f{5}{2}}_H$\\\hline
			Longitudinal thermal &&&&&\\ conductivity,$\kappa_{xx}=\kappa_{yy}$ &$T_H$&$T_H$&$T_H$&$T_H$& $T_H$\\\hline
			Transverse thermal &&&&&\\ conductivity, $\kappa_{xy}=-\kappa_{yx}$ &&$\rho B T^{0}_H$&-$ \rho B T^{0}_H$&$ -\rho B T^{-1}_H$& $T^{-1}_H$\\\hline
		\end{tabular}
	\end{center}\caption{\label{table}The precise temperature dependence of the transport  quantities  in $3+1$ spacetime dimension with magnetic field.} 
\end{table}

\section{Discussion}

It is suggested in \cite{Hartnoll:2015sea} that in order to study the scaling behavior of transport coefficients, one need to invoke some principles. It involves time reversal invariance and particle-hole symmetry. Some comments are in order.

\paragraph{Time reversal invariance:} Under time reversal operation, the current densities and the magnetic fields are odd whereas the electric fields and charge densities are even.  Then it follows trivially that the electrical conductivity, thermo-electric conductivity and thermal conductivity matrices are odd under time reversal operation. This explains nicely the linear dependence of magnetic field on the transverse electrical conductivity but  not on the longitudinal electrical conductivity. In fact, from the explicit  structure of the conductivities as written down in eq(\ref{gen_transport}), eq(\ref{gen_kappa}), eq(\ref{lifshitz_transport}) and eq(\ref{hsv_transport}) it follows that  none of the longitudinal conductivities are odd under time reversal symmetry.  

The basic reason behind such behavior of the longitudinal conductivities can be understood  as follows: If we look at the expression of the currents as written in eq(\ref{electrical_current}), eq(\ref{heat_current}), eq(\ref{electric_current_b}), eq(\ref{heat_current_qx}) and eq(\ref{heat_current_qy}) then they are odd under time reversal symmetry. However, the currents evaluated at the horizon eq(\ref{electric_current_horizon}) and eq(\ref{heat_current_horizon}) are not. This is  due to the in-falling boundary condition for the metric components and gauge potential at the horizon, which essentially breaks it. This explains the behavior of the longitudinal part of the conductivities under time reversal symmetry. How about the transverse part of the conductivities?

In order to understand the transverse part of the conductivities, let us look in detail at the behavior of the metric components at the horizon, in particular, $h_{tx}(r_h)$ and $h_{ty}(r_h)$. In the absence of  magnetic field  these metric components at the horizon breaks the time reversal symmetry. However, in the presence of a magnetic field, these  metric components possesses two parts. One part that breaks the time reversal symmetry and the other part that preserves it. From eq(\ref{htx_hty_horizon_b}), there follows\footnote{ when $U_1(r)=U_2(r)$, it is also shown in eq(26) of \cite{Pal:2019bfw}.} such a decomposition of the metric components at the horizon. It follows that one can write
\be
h_{ti}(r_h)=E_if(r_h)+\xi_i {\tilde f}(r_h)+{\epsilon_i}^jE_jg(r_h)+{\epsilon_i}^j\xi_j{\tilde g}(r_h),\quad {\rm where~~ i,~~j=x,~~y},
\ee

where the functions $f(r_h),~~ {\tilde f}(r_h),~~ g(r_h)$ and ${\tilde g}(r_h)$  are functions of  magnetic field, charge density, dissipation parameter and couplings etc. $\epsilon_{ij}$ is the Levi-Civita symbol. Importantly, it is the term involving $g(r_h),~~{\tilde g}(r_h)$, in the above equation are odd under  time reversal operation whereas the  term involving $f(r_h),~~{\tilde f}(r_h)$, are even under time reversal operation. 

The current when evaluated at the horizon has got two pieces. One piece that is even under the time reversal symmetry and the other is odd. Upon writing down the
current as
 \be
 J^{i}(r_h)=\delta^{ij}E_j f_1(r_h)+\delta^{ij}\xi_j  f_2(r_h)+\epsilon^{ij}E_j g_1(r_h)+\epsilon^{ij}\xi_j g_2(r_h),\quad {\rm where~~ i,~~j=x,~~y},
 \ee
where $f_i$'s and $g_i$'s  are functions of the magnetic field, charge density, dissipation parameter and couplings etc. The functions $f_1$ and $f_2$ are even under time reversal whereas   $g_1$ and $g_2$ are odd.
The functions $g_1$ and $g_2$   contributes to the transverse conductivities whereas the functions  $f_1$ and $f_2$  contributes to the longitudinal conductivities. Similarly, for the heat current
\be
{\cal Q}^{i}(r_h)=\delta^{ij}E_j {\tilde f}_1(r_h)+\delta^{ij}\xi_j   {\tilde f}_2(r_h)+\epsilon^{ij}E_j  {\tilde g}_1(r_h)+\epsilon^{ij}\xi_j  {\tilde g}_2(r_h),\quad {\rm where~~ i,~~j=x,~~y},
\ee
where ${\tilde f}_i$'s and ${\tilde g}_i$'s  are functions of the magnetic field, charge density, dissipation parameter and couplings etc. Once again the functions  ${\tilde f}_i$'s and ${\tilde g}_i$'s are even and odd under time reversal symmetry, respectively. This explains the necessary structure of the longitudinal and transverse conductivities under time reversal operation. 

To summarize, the transport quantities are calculated at the horizon and at the horizon the in-falling boundary conditions breaks the time reversal symmetry. 

\paragraph{Particle-Hole symmetry:} It is suggested in \cite{Hartnoll:2015sea} that the theory around the quantum critical point is not particle-hole symmetric.  In our study of  the transport quantities and others it is found that some of the transport quantities are symmetric under particle-hole symmetry while some are not.

As argued in \cite{Donos:2014cya, Blake:2014yla} in the absence of magnetic field the longitudinal conductivity consists of two terms. One term with the dissipation due to the lattice and the other is without the dissipation.  In \cite{Blake:2014yla}, the authors have suggested that it is the particle-hole symmetric term, which is   without the dissipation,  contributes to the linear in temperature to resistivity and it is this term that dominates over the dissipative term.  In contrast to this, it is reported in \cite{Davison:2013txa} that it is the dissipative term that contributes to the linear in temperature to resistivity. In this paper, we show in fact  that both the terms contribute to the resistivity having the same temperature dependence.

The future work would be to construct models which will show the temperature dependence of all the transport quantities as has been observed experimentally for strange metal.

\paragraph{Acknowledgment:} It is a pleasure to thank Sean Hartnoll for going through the draft and making several suggestions. Thanks are to  arxiv.org, gmail  and inspirehep.net for providing their support throughout this work.

\section{Appendix A}

\paragraph{Conductivity for Maxwell system:} Starting from the electric current as written in eq(\ref{maxwell_current}), the x-component of it reads as
\be
J^x_{Maxwell}(r)=-Z(\phi(r)) \sqrt{-g} F^{rx}.
\ee
Let us consider the following $d+1$ dimensional spacetime
\be
ds^2=-U_1(r)dt^2+h(r)\sum^{d-1}_{i=1}dx^idx_i+\f{dr^2}{U_2(r)}.
\ee
By keeping the x-component of the fluctuated gauge field as $a_x(r)$, we find the current reads as
\be
J^x_{Maxwell}(r)=-Z(\phi(r))\sqrt{U_1(r)U_2(r)}h^{\f{d-3}{2}}(r)a'_x(r).
\ee 
With the help of in-falling boundary condition on the fluctuated gauge field $a_x(r)$ and the  behavior of $U_1(r)$ and $U_2(r)$ at the horizon gives the current as
\be
J^x_{Maxwell}(r_h)=Z(\phi(r_h))h^{\f{d-3}{2}}(r_h)E_x.
\ee
The desired longitudinal electrical conductivity eq(\ref{cond_maxwell}) follows for the hyperscale violating solution, i.e., we have set $h(r)=r^{2(1-\gamma)}$.

\paragraph{Solution in the limit of small magnetic field of eq(\ref{diff_eq_b}) and eq(\ref{scalar_diff_eq_ir_b}): } To quadratic order in the magnetic field, the  equations are 
 
\bea
&&2U_2 U''_1+U'_1 U'_2-\f{U_2 U'^2_1}{U_1}+\f{2U_2 h'U'_1}{h}+2U_1(2\Lambda+V)+2T_b \f{ U_1h Z^2_1Z^3_2}{\sqrt{\rho^2+h^2Z^2_1Z^2_2}}\nn&&-T_b\f{U_1h Z^4_1Z^3_2 B^2}{(\rho^2+h^2Z^2_1Z^2_2)^{3/2}}=0,\nn
&&2 U_2 h''+h' U'_2+\f{U_2 h' U'_1}{U_1}+2k^2\psi+2h(2\Lambda+V)+2T_bZ_2\sqrt{\rho^2+h^2Z^2_1Z^2_2}+T_b\f{Z^2_1Z_2B^2}{\sqrt{\rho^2+h^2Z^2_1Z^2_2}} =0,\nn
&&2h^2U_1U_2 U''_1+4hU^2_1U_2 h''+2h^2U^2_1(2\Lambda+V+U_2\phi'^2)+U'_2hU_1(hU'_1+2U_1h')-\nn&&U_2(h^2U'^2_1+2U^2_1h'^2)+2T_b \f{h^3 U^{2}_1Z^2_1Z^3_2}{\sqrt{\rho^2+h^2Z^2_1Z^2_2}}-T_b \f{h^3 U^2_1Z^4_1Z^3_2 B^2}{(\rho^2+h^2Z^2_1Z^2_2)^{3/2}}=0, \nn
&&h\phi''+\f{(2U_1U_2h'+hU_2U'_1+hU_1U'_2)}{2U_1U_2}\phi'-
\left(\f{h}{U_2}\f{dV}{d\phi}+\f{k^2}{U_2}\f{d\psi}{d\phi}\right)-\nn&&T_b\f{(h^2Z_1Z^3_2Z'_1+Z'_2(\rho^2+2h^2Z^2_!Z^2_2))}{U_2\sqrt{\rho^2+h^2Z^2_1Z^2_2}}-T_b B^2Z_1\f{(\rho^2Z_1Z'_2+Z_2Z'_1(2\rho^2+h^2Z^2_1Z^2_2))}{2U_2(\rho^2+h^2Z^2_1Z^2_2)^{3/2}} =0,\nn
\eea
where we have used eq(\ref{gauge_field_ir_b}) in eq(\ref{diff_eq_b}) and eq(\ref{scalar_diff_eq_ir_b}). In order to solve the equation we consider logarithmic behavior of the dilaton. 
\bea
\phi(r)&=&\varphi_0 Log~r,\quad Z_1(\phi(r))=e^{-\alpha\phi(r)}=r^{-\alpha\varphi_0},\quad Z_2(\phi(r))=e^{\beta\phi(r)}=r^{\beta\varphi_0},\nn
\psi(\phi(r))&=&\psi_0 e^{\psi_1\phi(r)}=\psi_0r^{\psi_1\varphi_0},\quad V(\phi(r))=m_1e^{m_2\phi(r)}=m_1r^{m_2\varphi_0}
\eea
The metric components are assumed to take the following form
\be
U_1(r)=r^{2(z-\gamma)}f(r),\quad U_2(r)=r^{2(1+\gamma)}f(r),\quad h(r)=r^{2(1-\gamma)}.
\ee
The above differential equations for $\Lambda=0$ are solved for
\bea
\alpha&=&\f{4-2\gamma}{\varphi_0},\quad \beta=\f{-2+2\gamma+\alpha\varphi_0}{\varphi_0},\quad \psi_1=-\f{2-2\gamma+\alpha\varphi_0}{\varphi_0},\nn m_2&=&-\f{4-4\gamma+\alpha\varphi_0}{\varphi_0},\quad \psi_0=-\f{ B^2(1+z-2\gamma)(2+z-2\gamma)(z-\gamma)}{k^2(\gamma-1)},\nn
m_1&=&\f{ B^2(2+z-2\gamma)(z-\gamma)^2}{\gamma-1},\quad \varphi_0=2\sqrt{(\gamma-1)(1-z+\gamma)},\quad f(r)=1-r^{-2-z+2\gamma}.\nn
\eea
We shall get a consistent solution provided, we set $T_b=2 (m_1/B) (1+\rho^2)^{3/2}$ and the charge density as $\rho=\sqrt{z-1}/\sqrt{\gamma-z}$.

\end{document}